# Electric-Field Control of Chirality

Piush Behera[1,2, †], Molly A. May[3, †], Fernando Gómez-Ortiz[4, †], Sandhya Susarla[2], Sujit Das[1], Christopher T. Nelson[5], Lucas Caretta[1], Shang-Lin Hsu[1,2], Margaret R. McCarter[2,6], Benjamin H. Savitzky[7], Edward S. Barnard[8], Archana Raja[8], Zijian Hong[9], Pablo García-Fernandez[4], Stephen W. Lovesey[10], Gerrit van der Laan[10], Colin Ophus[7], Lane W. Martin[1,2], Javier Junquera[4,*], Markus B. Raschke[3], Ramamoorthy Ramesh[1,2,6,*]

[1]Department of Materials Science & Engineering, University of California; Berkeley, CA 94720.

[2]Materials Sciences Division, Lawrence Berkeley National Laboratory; Berkeley, CA 94720.

[3]Department of Physics, Department of Chemistry and JILA, University of Colorado; Boulder, CO 80309.

[4]Departamento de Ciencias de la Tierra y Física de la Materia Condensada, Universidad de Cantabria; Cantabria Campus Internacional 39005 Santander, Spain.

[5]Center for Nanophase Materials Sciences, Oak Ridge National Laboratory; Oak Ridge, TN, 37831.

[6]Department of Physics, University of California; Berkeley CA 94720.

[7]National Center for Electron Microscopy, Lawrence Berkeley National Laboratory; Berkeley, CA, 94720.

[8]The Molecular Foundry, Lawrence Berkeley National Laboratory; Berkeley, CA, 94720.

[9]Department of Mechanical Engineering, Carnegie Mellon University; Pittsburgh, PA, 15213.

[10]Diamond Light Source, Harwell Science and Innovation Campus; Didcot, Oxfordshire OX11 0DE, United Kingdom.

† These authors contributed equally
*Corresponding Author.  E-mail: *rramesh@berkeley.edu*; *javier.junquera@unican.es*

**Abstract:** Polar textures have attracted significant attention in recent years as a promising analog to spin-based textures in ferromagnets. Here, using optical second harmonic generation based circular dichroism, we demonstrate deterministic and reversible control of chirality over mesoscale regions in ferroelectric vortices using an applied electric field. The microscopic origins of the chirality, the pathway during the switching, and the mechanism for electric-field control are described theoretically via phase-field modeling and second-principles simulations, and experimentally by examination of the microscopic response of the vortices under an applied field. The emergence of chirality from the combination of non-chiral materials and subsequent control of the handedness with an electric field has far-reaching implications for new electronics based on chirality as a field controllable order parameter.

Of all the fundamental symmetries observed in nature and particularly in solid-state systems (*e.g.*, time-reversal and spatial-inversion symmetries and the combination of these two through a toroidal order[1,2]) chirality is perhaps the most exotic and yet pervasive. It can be defined as an asymmetric configurational property where an object cannot be superimposed on its mirror image, thereby imparting a handedness. Chirality is a basic feature that determines many important properties in nature, from the strength of the weak interactions according to the electroweak theory, to its essential role in the spontaneous symmetry breaking in subatomic particle physics, or biophysics.[3,4] A prime example of this are the building blocks of life itself, which is built up from molecules that are exclusively *L*-amino acids and can only convert *d*-glucose as a fundamental source of energy. Enantiomeric conversion of sugars such as *d*-glucose to *l*-glucose is typically accomplished by chemical means and is inherently a destructive process (the molecule is broken and reformed anew)[4,5]. Chirality appears in inorganic systems as well, either as a consequence of the electronic spin structure (*e.g.*, ferromagnets) or mesoscale arrangements of molecular building blocks (*e.g.*, liquid crystals)[6,7]. Furthermore, handed responses to optical stimuli can be designed into engineered structures, such as the observation of the photonic spin-Hall effect in designer metamaterials[8,9]. Due to the universal nature of chirality, manipulation of the handedness by thermodynamic fields other than chemical fields would be of great interest, both scientifically and technologically with the potential for chiral electronics[10].

Over the past several years, non-trivial topological dipolar textures have emerged as an area of great attention in condensed-matter research[11–13], building from and complementing the on-going revolution in topological spin-based textures such as chiral skyrmions and vortices[14–17]. Such dipolar textures have been mainly identified in epitaxial heterostructures in which exact electrostatic and elastic boundary conditions are imposed on a crystalline ferroelectric layer, such as $PbTiO_3$[13,18,19]. Dipolar textures in ferroelectrics, such as polar vortices and skyrmions have been extensively studied and several interesting physical phenomena have been discovered, such as toroidal order and negative capacitance[20–22]. Perhaps the most exotic and unexpected among them is the emergence of chirality (*i.e.*, handedness) in polar vortices and skyrmions that arise from a sequence of achiral materials[20,23]. The existence of chiral behavior in such systems was demonstrated using resonant soft X-ray diffraction based circular dichroism (RSXD-CD) measurements[20,24,25]. However, the use of such synchrotron-based approaches, while scientifically sophisticated, can limit the pervasive impact of such chiral textures, particularly for possible applications in next generation technologies. In parallel, strain engineering or electric-field manipulation and control of chiral phases would be of fundamental and technological interest. This has been the subject of theoretical studies[26–28] but has remained elusive from the experimental point of view. A recent work addresses this problem, but the presence of random fields during the switching precludes a deterministic control of the vortex rotation reversal[29]. - With this as background, the present work introduces two main innovations: (i) the manipulation of the chirality in a controlled, deterministic, and reversible way in polar vortices over mesoscale regions using an electric field; and (ii) the use of conventional optical techniques to probe chirality in such polar textures.

For this work, symmetric $(SrTiO_3)_{20}/(PbTiO_3)_{20}/(SrTiO_3)_{20}$ (STO/PTO/STO) trilayers were synthesized on orthorhombic $DyScO_3$ (DSO) (110) substrates using reflection high-energy electron diffraction (RHEED)-assisted pulsed-laser deposition (Methods and Ref. [13]). In such heterostructures, the vortex state dominates due to the interplay between depolarization energy at the PTO/STO interfaces, elastic energy from the tensile strain imposed by the DSO substrate, and the gradient energy in the ferroelectric[13,20]. High angle annular dark field scanning transmission electron microscopy (HAADF-STEM) in conjunction with vector displacement mapping was used to visualize the local ion displacements in this vortex phase (Fig. 1a), which reveal vortices with alternating clockwise (CW)/counterclockwise(CCW) polarization curls in the PbTiO$_3$ layer. The cores of these neighboring vortices, indicated by white circles, are displaced vertically in opposite directions from the mid-point (indicated by white crosses in Fig. 1a). As a consequence, this buckling (*i.e.,* a staggered vortex configuration) leads to a net in-plane polarization along the $[001]_o$, which is validated by phase-field modeling (Fig. 1b; details of the phase-field modeling are provided in the Methods).

Chirality can arise in ensembles of such polar vortices as a consequence of a few symmetry-breaking pathways, which are schematically illustrated (Figs. 1c,d,e). As has been previously shown[20], chiral behavior can arise due to the coexistence of the axial component of polarization (perpendicular to the plane defined by the vortices, here the $[1\bar{1}0]_o$) with the vorticity of the CW and CCW vortices[23]. Mathematically, it is described by the helicity $\mathcal{H}$, defined as

$$\mathcal{H} = \int \vec{p} \cdot (\vec{\nabla} \times \vec{p}) \, d^3r, \qquad (1)$$

where $\vec{p}$ is the local value of polarization, thus leading to either a left-handed (negative value of $\mathcal{H}$) or right-handed (positive $\mathcal{H}$) vortex array that is controlled by the axial polarization direction, as schematically described in Fig. 1c. How the lateral and axial components of the polarization are measured is described in the Methods and represented below. A second source of chirality can be observed when the vortices are offset along the $[110]_o$, normal to the vortex axis (Fig. 1d). These misalignments lead to a mismatch in the polarization pointing along the

$[001]_o$

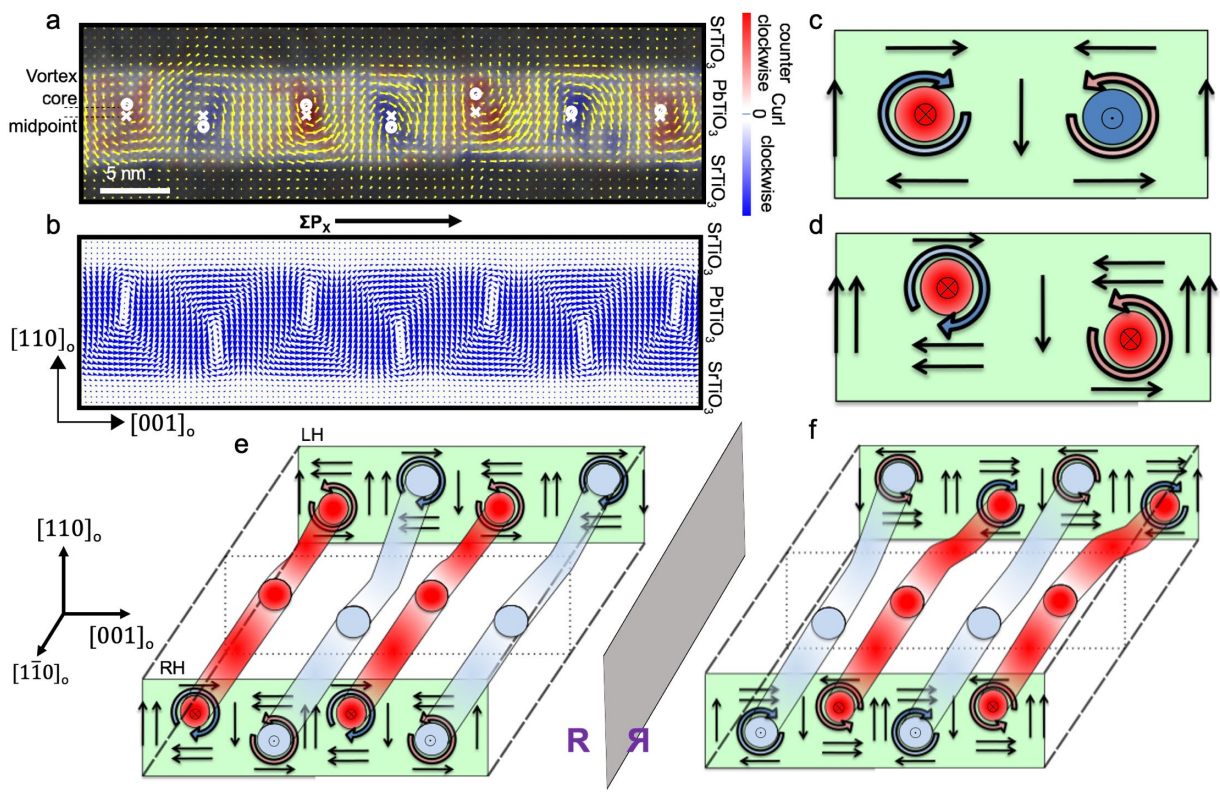

**Figure 1: Different mechanisms for achieving chirality. (A)** HAADF-STEM displacement map of polar vortex array. Color indicates the curl of the local polarization. Circles represent the core of the vortices and crosses the center points of the PbTiO₃ layer. **(B)** Phase field modeling of vortex array demonstrating pronounced buckling of vortices. **(C)** CW (blue curled arrow) and CCW (orange curled arrow) vortices coupled with antiparallel axial components of the polarization. Filled red and blue circles represent the direction of the axial polarization (along the $[1\bar{1}0]_o$ direction), light colors represent lower values. Straight arrows represent the direction of the local polarization along the $[001]_o$ (horizontal) or $[110]_o$ (vertical) directions. Within the right-hand rule discussed in Ref. [20], both vortices in the sketch are right-handed. **(D)** Asymmetry between up and down polarized domains (along $[110]_o$), superimposed on an offset that generates a mismatch between left and right polarization along the $[001]_o$ direction. In the sketch, at the center of the ferroelectric layer, the two vortices point predominantly to the left. The handedness of the two vortices is opposite, but one of them is slightly larger than the other, giving the whole system a net chirality. **(E)** Schematic representation of the right-handed and left-handed domains experimentally observed, separated by a domain wall (dashed square at the center). The two sources of chirality (antiparallel axial components at the center of consecutive vortices plus an offset of its center) coexist within a domain. At the domain wall, the sense of rotation of a vortex is reversed, keeping constant the sense of the axial polarization. **(F)** Vortices of **e** after performing a mirror symmetry operation, represented by the "R". Schematics with the three orthogonal reflections of the polar texture supercell to determine their chiral nature are presented in Fig. S1.

(left/right). If this offset appears in combination with a non-equal fraction of dipoles pointing along the vertical $[110]_o$ (i.e., if the up and down domains do not exactly equally match in size), then an excess/deficit of the CW/CCW rotations is generated, making the whole system chiral (see Fig. S2). Therefore, for given sizes of the up and down domains, the structure exhibits a chirality that is dependent on the direction of the mismatch (or buckling). From a quantitative point of view, the degree of chirality can be captured by the helicity defined earlier. While the sign of the helicity allows an unambiguous classification of the handedness of a given array of polar vortices, its absolute magnitude determines the strength of the chiral behavior (*e.g.*, the magnitude of circular dichroism; see Supplementary Section D for details of the strength of the chirality). The second pathway (Fig. 1d) produces helicities that are smaller in magnitude than those of the first (Fig. 1c) and arises from an imbalance in the center of rotation of the two vortices. This pathway has the advantage, however, that it is potentially easier to deterministically reverse the vortex offset with an external electric field applied perpendicular to the lateral component of the polarization (Fig. S1b) and thus the net left/right polarization direction. Indeed, the different paths for switching the chirality of the model depicted in Fig. 1c are energetically costly or impractical (Fig. S3). Experimentally, we observe vortex structures that have attributes of both sources of chirality (Figs. 1c,d), *i.e.*, an antiparallel axial component that is superimposed on an up/down shift of the CW/CCW rotating vortices. Indeed, experimentally, domains with different chirality separated by a thin domain wall are observed, as schematically shown in Fig. 1e. At the domain wall, the rotation of a given tube is reversed, with a corresponding change in the position of the vortex core, keeping constant the axial component of the polarization. This domain structure is chiral, since it cannot be superimposed to any mirror image (Fig. 1f and Fig. S1c). Moreover, reversal of the net polarization along the $[001]_o$ results in a reversal of the buckling pattern, providing a possible pathway to switch the chirality with an electric field. We demonstrate such an electric field driven reversal of the chirality using optical second harmonic generation based circular dichroism (SHG-CD) as the indicator of chirality (Methods) and validated by electric-field dependent electron microscopy studies.

The SHG-CD studies were carried out in a confocal microscope with spatial resolution of ~300 nm as a convenient, laboratory-based approach to study chirality (Methods and Supplementary Section E). SHG is a photonic exchange between the frequency components of the electromagnetic field, during which two photons of lower frequency $\omega$, are absorbed and one photon of $2\omega$ is created in a single quantum-mechanical process. SHG fundamentally arises from symmetry breaking in the second-order susceptibility, making it a powerful tool for imaging crystal lattices with symmetry breaking, such as ferroelectrics[30]. Moreover, CD in SHG is an established technique to probe chirality[31–34] and can result from, for instance, chiral ordering of electric dipoles or the existence of ferromagnetism[35,36]. The lack of magnetism in PTO and STO suggests that any SHG-CD that arises must come from a chiral ordering of dipoles. Such a natural circular dichroism (NCD) of non-magnetic origin can only arise from the parity-odd event of three electric-dipole transitions. Though all non-centrosymmetric materials can display such a transition, only when helical light interacts with a handed environment will the magnitude of the NCD be non-zero. (Details of the full derivation are provided in the Supplementary Section G)[35].

Confocal scanning microscopy images of the SHG signal with right- (RC) and left- (LC) circularly polarized excitation are shown (Figs. 2a and b, respectively). The SHG-CD signal is then calculated from the asymmetry of intensity distribution between the two images,

$$\frac{I_{LC} - I_{RC}}{I_{LC} + I_{RC}} \qquad (2)$$

where $I_{LC}$ and $I_{RC}$ are the SHG intensities when excited by a LC and RC polarized beam, respectively. The resulting CD image (Fig. 2c) reveals regions with strong SHG-CD signals (magnitude as large as 0.4), strongly pointing to the existence of chiral ordering. Notably, domains with opposite CD elongated along the $[001]_o$ are visible, indicating domains or regions of vortices with opposite chirality. It is worth noting that while the characteristic size of the domains imaged in SHG-CD is ~1 μm, this is close to the diffraction-limited spatial resolution of the microscope (~300 nm with an oil immersion lens) and does not rule out the existence of features below this length scale.

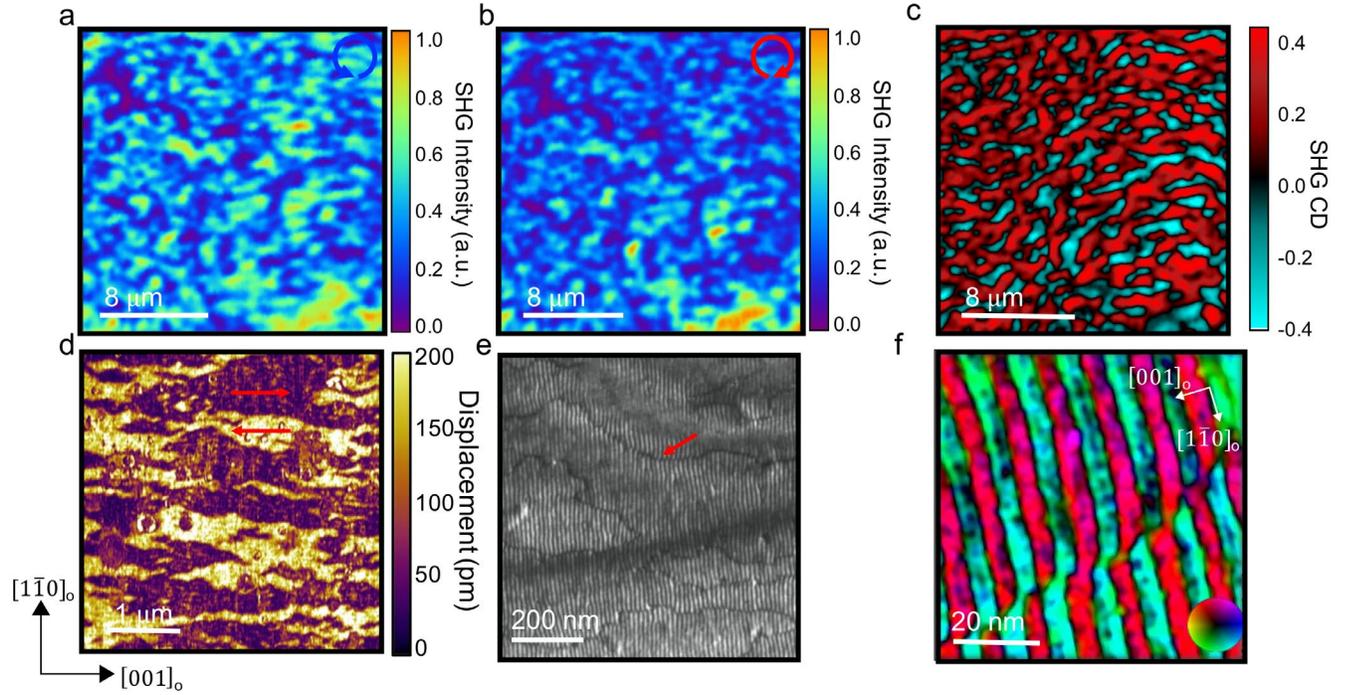

**Figure 2: Vortex Domain Structure.** SHG images taken with **(A)** right-circularly (RC) and **(B),** left-circularly (LC) polarized excitation. **(C)** SHG CD calculated from the images shown in **a** and **b** using equation 2. **(D)** Lateral PFM image with red arrows indicating regions of opposite polarization along $[001]_o$. **(E)** Weak beam DF-TEM image of vortex domains. Arrow shows location of antiphase domain wall **(F)** HAADF-STEM displacement map at a vortex domain wall. Color indicates direction of atomic displacement with respect to the color wheel.

In-plane piezoforce microscopy (PFM) studies reveal a lenticular-shaped domain pattern that is captured in Fig. 2d with anti-parallel in-plane polarization components, identified by the red arrows. The lateral length scales of such domains are in the 100-300 nm range while they extend to ~1 μm in the longitudinal direction ($[001]_o$). The ferroelectric domains are qualitatively similar in morphology to the chiral domains in the SHG-CD image and suggest a possible causal dependency between them. At higher resolution, for example in a weak beam dark field TEM image (DF-TEM), Fig. 2e, we are able to observe the arrays of vortices with their long axis along the $[1\bar{1}0]_o$ (vertically in Fig. 2e). More interestingly, we observe contrast arising from antiphase domain walls, indicated by the red arrow. At higher resolution, for instance using a HAADF-STEM based atomic imaging, we can discern the displacement maps of the atomic structure across such interfaces, as illustrated in Fig. 2f.

We next focus on the relationship between the lateral and axial components of the polar structural distortion across such boundaries. Since HAADF-STEM is mainly a 2D projection technique, it cannot deterministically probe the axial component. Instead, we used four dimensional (4D)-STEM to probe the in-plane and axial polarization components. In 4D-STEM, a focused probe is scanned across the sample region to obtain a convergent beam electron diffraction (CBED) pattern at every scan position. Unlike HAADF-STEM, 4D-STEM has a depth dependence, providing additional information through the thickness of the sample[37]. Due to this advantage it is easier to measure and distinguish between the lateral and axial polarization. Fig. 3a shows the HAADF image where the 4D-STEM was performed and Fig. 3b shows the CBED pattern averaged over the entire image in Fig. 3a. To study the change in chirality across a domain boundary (white dashed line in Fig. 3a), we obtained polarization maps by subtracting the normalized intensity of the virtual images formed from the Friedel pair CBED disks along the lateral (disk 1 and disk 2) and axial (disk 3 and disk 4) directions (See Fig. S6). Polarization maps of the lateral and axial component, Fig. 3c and Fig. 3d respectively, were generated from within the orange boxed region in Fig. 3a. Further details of this 4D-STEM imaging are described in the Methods and Supplementary Section H. From such images, we obtain the line scans from the two regions across the domain boundary (Fig. 3e,f). Focusing on the region outline by the solid black line in Fig. 3c,d, we see that the lateral and axial components are in-phase with one another, as shown in Fig. 3e. In contrast, in the region outlined by the dotted black lines the lateral and axial components are out-of-phase with each other. Such a change in the phase in the lateral and axial components has been shown to be an indicator of the change in the chirality[38].

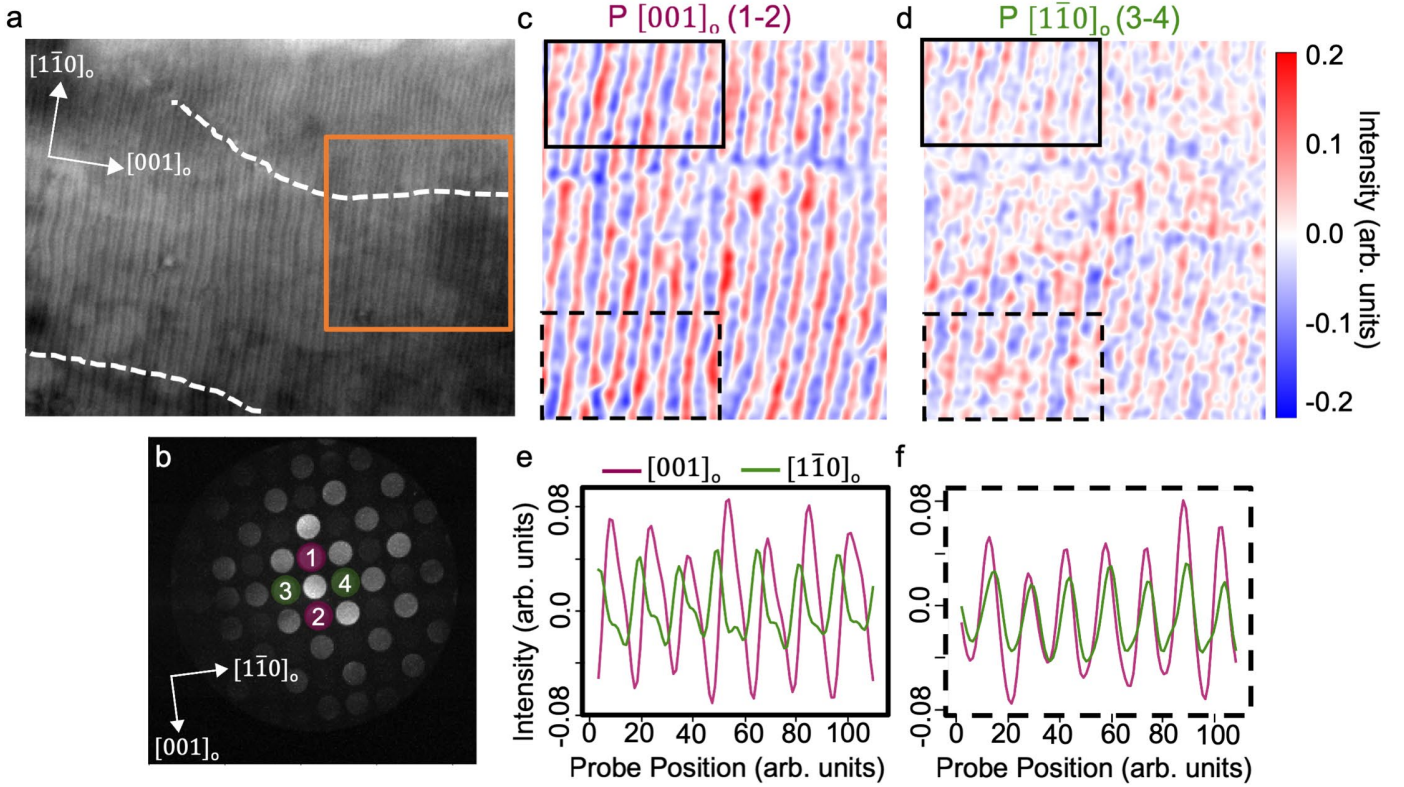

**Figure 3: 4D STEM-analysis of in-plane vortices. (A)** Large area HAADF-STEM showing the in-plane vortices along $[1\bar{1}0]_o$ direction with domain boundaries indicated by the grey dotted line. **(B)** A representative CBED pattern from the vortex structure shown in (a), which identifies the reflections that were used to form the 4D-STEM images in c,d. **(C)** Lateral and **(D)** axial polarization components extracted from orange square region indicated in a. **(E,F)** Line scans from different regions across the domains, that show lateral and axial polarization out-of-phase and in-phase respectively.

These results provide the ingredients that support the chiral and switchable atomic model schematized in Fig. 1e. In particular, (i) the observation of alternating CW and CCW vortices running along the $[1\bar{1}0]_o$ direction, visualized in the red and blue stripes for the $[001]_o$ component of the polarization in Fig. 3b; (ii) the existence of antiparallel axial components of the polarization along the $[1\bar{1}0]_o$ axial direction in consecutive vortices, as shown also by the red and blue stripes in Fig. 3c; (iii) The pronounced buckling of the vortices observed by HAADF-STEM in Fig. 1a; and, (iv) the presence of domains of different helicity, separated by the domain wall marked with the dotted line Fig. 3a-c. The character of the domains is clearly visible in Figs. 3d-e, where the components of the polarization along the $[001]_o$ and $[1\bar{1}0]_o$ are, respectively, in-phase and out-of-phase, indicating a change in the handedness, as described schematically in Fig. 1e.

Having established the equilibrium chiral domain structure, we next turn to their *in situ* manipulation during the SHG-CD measurements. A DC electric field was applied along the in-plane $[001]_o$ using lithographically patterned interdigitated electrodes (Fig. S7a,b). This pattern leads to linear electrostatic potentials, *i.e.*, constant electric fields of opposite polarity for adjacent interdigitated regions (as indicated by the white arrows, Figs. 4a,b). The direction of the DC field favors the $[001]_o$ component of the polarization in the same sense and, therefore, selects the direction of the buckling of the vortices. This offset of the vortex cores determines unequivocally the helicity, as shown in Fig.1e-f, and supported by the second-principles simulations in Fig. S10. Therefore, the chirality can be deterministically controlled by the external field. An SHG-CD image taken with an applied field of ±50 kV/cm (±40 V across an 8 micron electrode spacing) within neighboring interdigitated regions is shown (Figs. 4a,b). The applied electric field leads to the formation of large bands of uniform CD across the entire region between the interdigitated electrodes, with the sign of the CD response being the measure of the sign of the chirality. As expected, reversing the sign of the applied voltage switches the sign of the CD (chirality) in each region between the electrodes. The robustness of the switching is illustrated by the line profiles (Figs. 4c,d), which correspond to the SHG-CD along the white, dashed lines (Figs. 4a,b) for a sequence of opposite polarity fields. The SHG-CD in these saturated regions is also larger in magnitude (by ~50%) than that observed in the unmodified samples. This increased magnitude likely arises due to two factors. First, the average size of polar domains in the virgin sample is on the order or even smaller than the diffraction-limited imaging resolution of the SHG measurements, which leads to averaging over regions with positive and negative CD responses, consequently decreasing the magnitude of the resultant CD.

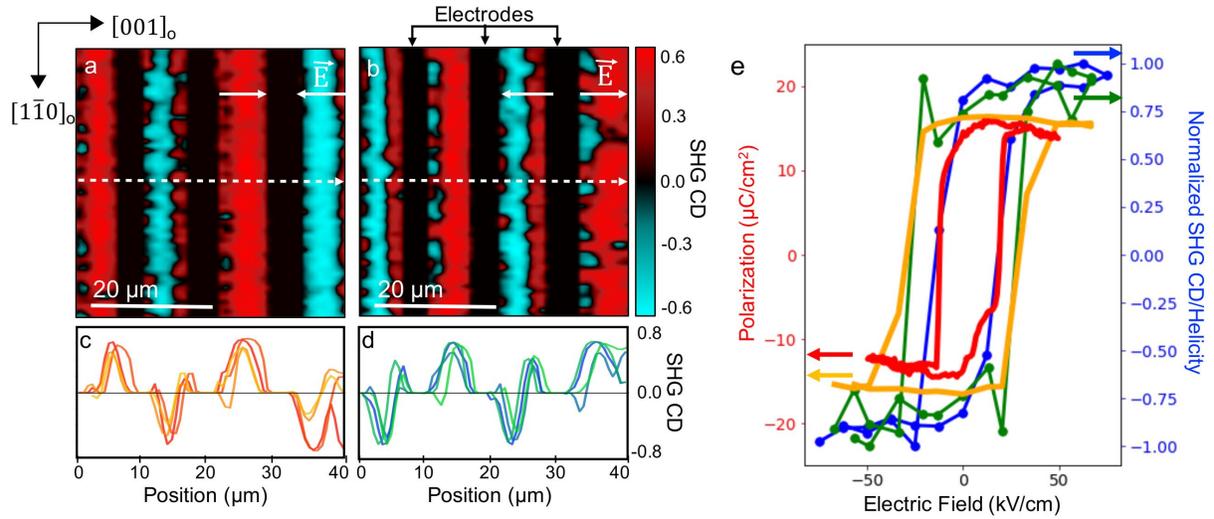

**Figure 4: Electric Field Switching of Chirality. (A,B)** SHG CD for alternating applied electric fields of ±50 kV/cm as indicated by the white arrows. **(B)** SHG CD for the same region shown in **a** but with inverted electric field polarities. **(C,D)** Data along dashed white lines in (A) and (B) with repeated reversal of the electric field polarity. Subsequent scans shown going from red/blue to orange/green for (A/B). **(E)** Hysteresis of the chiral switching measured using SHG CD (blue) and calculated helicity (green),both of which are normalized. Polarization measured from an electric field applied along the $[001]_o$ in-plane direction (red), and calculated polarization hysteresis curve from second principles (orange).

Second, the application of an electric field induces larger offsets between the core of the vortices and, therefore, there is an enhancement of the helicity.

The DC voltage dependence of the SHG-CD measurements (Fig. 4e, blue data) reveals a CCW, hysteretic nature of the field dependence of the normalized SHG-CD (chirality) with a coercive field of ~20 kV/cm and a nonvolatile remanent chiral state. This is a direct indication of the switchability of chirality by an electric field. Data illustrating reproducibility of the hysteresis effect is also presented in Fig. S8a,b. To probe this further, we carried out polarization-electric field hysteresis measurements using the same interdigitated electrodes to probe the switching of the in-plane component of the switchable polarization (Fig. 4e, red data). The raw data for this ferroelectric hysteresis loop measurement (Fig. S9) shows the existence of a linear dielectric component superimposed on the ferroelectric component. For the sake of simplicity, we have subtracted out the linear dielectric component (Fig. 4e). The overlap of the ferroelectric polarization hysteresis loops with the SHG-CD hysteresis, establishes a direct correlation between the in-plane polarization (arising from the buckling pattern in the vortex arrays) and the consequent chiral behavior as discerned from the SHG-CD measurements.

In order to shed light on the microscopic origin of the switchable chirality, particularly the change in the sense of buckling of the vortices, we have carried out second-principles, simulations[39,40] on $(PbTiO_3)_{10}/(SrTiO_3)_{10}$ superlattices, using the same lattice potential as in Ref. [20] (see Methods). The polarization textures obtained after the relaxation of the atomic structures are similar to those previously reported[20], *i.e.*, the local dipoles display pairs of CW and CCW vortices alternating along the $[001]_o$ axis. The imposed biaxial tensile strain favors the development of a net in-plane polarization superimposed on the previous structure. As a consequence, and in order to accommodate the onset of this polarization, neighboring vortices arrange in a buckled geometry. Two enantiomorphic structures are possible (Fig. S10a,b), where it is seen how the direction of the in-plane polarization unequivocally determines the offset between the core of the vortices. A polarization pointing along the positive (negative) $[001]_o$ forces the center of the CW vortex to be lower (higher) than the core of the neighboring CCW vortex (Figs. 1e,f).

The ferroelectric hysteresis loops for the net in-plane polarization and the normalized integrated helicity are compared in Fig. 4e, (orange and green) respectively. When the external electric field points in the same direction as the net in-plane polarization, the offset between the vortices increases. This tendency is maintained until the core of the two vortices touch the PTO/STO interface (Fig. S10c,d). Beyond these fields, the vortices are destroyed, transforming themselves first in sinusoidal (wave-like) textures, and finally into achiral, monodomain configurations (zero helicity) with all the polarization pointing in-plane. The opposite happens when the external electric field points in the opposite direction to that of the in-plane polarization. In that case, the in-plane component of the polarization is progressively reduced, with the concomitant reduction of the offset (Fig. S10e-f). Beyond the coercive field, the in-plane polarization is reversed, followed by the direction of the offsets and the integrated helicity. The theoretically calculated hysteresis in the polarization, with the linear dielectric component subtracted, and helicity of the vortices (Fig. 4e), are in close agreement with the experimental data, with the coercive field being slightly larger at ~25 kV/cm. Polarization hysteresis raw data are also provided (Fig. S9)

We then probed the microscopic origins of this correlation between the buckling of the vortices and chirality as well as the electric-field manipulation of the buckling/chirality using cross-section DF-TEM studies. Dark-field images were obtained using the $[110]_o$ and the $[001]_o$ reflections (details of the full diffraction analyses and the dark-field images are presented in Fig. S11). From the combination of the diffraction condition and the image contrast, we are able to assign vectorial directions to the intensity maps (Fig. 5a); the contrast in the DF-TEM distinguishes regions of opposite polarization along the out-of-plane $[110]_o$. The arrows show the direction of the polarization within each region, and the magnitude of the arrows represent the area of each region. A characteristic chevron shaped pattern can be seen, and the intersections of the four regions represents the core of the vortex (red dots, Figs. 5a-c). *In situ* electric field studies were carried out (Fig. S13) on such a vortex array. With increasing electric field, the neighboring vortices flip the sense of their in-plane component at a voltage of ~5.6V. This is schematically described by the red circles and yellow arrows (Figs. 5a-c). This flipping (Fig. 5b) is a reversal of the buckling pattern described theoretically in Figs. 1b-d, and is indicative of a change in chirality, as inferred from the model (Figs. 1e,f). Upon reversing the voltage (Fig. 5c), the buckling pattern of the vortices is flipped back to the original state. Phase-field simulations that recreate the structures produced in the *in situ* TEM data (Fig. 5d,f) illustrate the switching of the vortex buckling under the application of an electric field along the $[001]_o$.

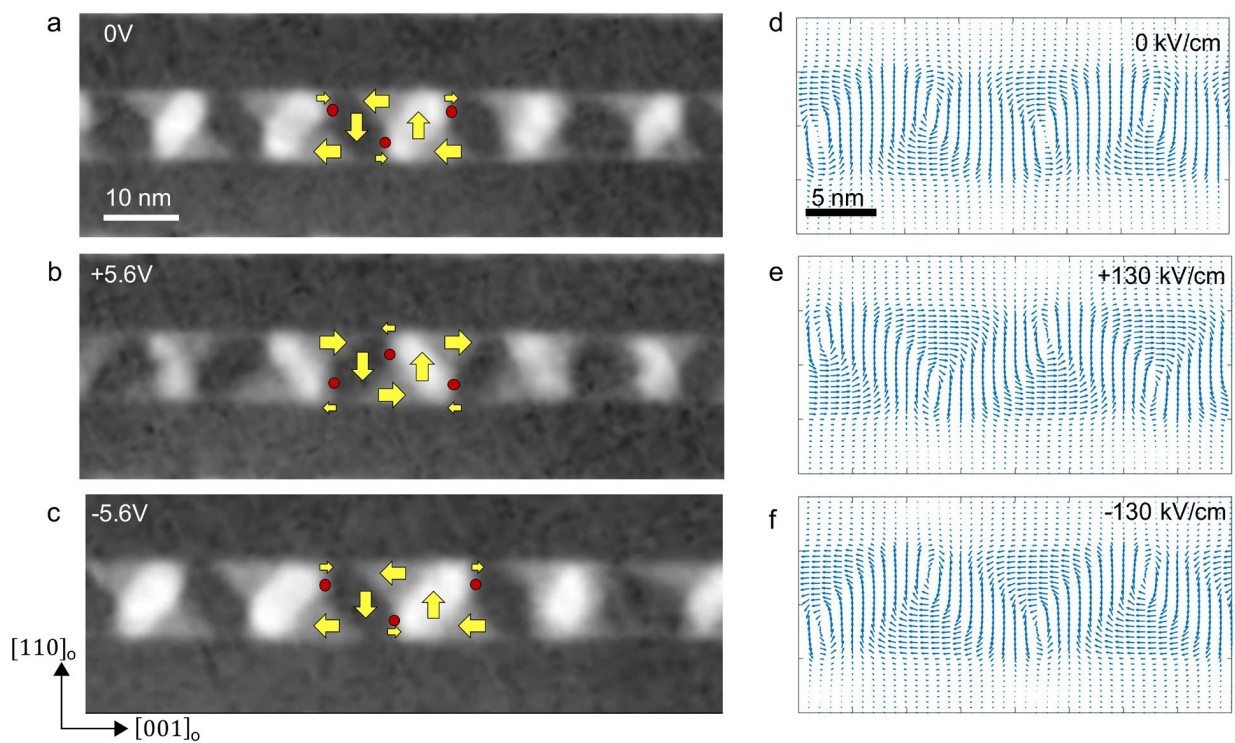

**Figure 5: Reversal of Vortex Buckling. (A-C)** DF-TEM showing atomic scale restructuring during chiral phase switch. Yellow arrows indicate direction of polarization and red circles indicate vortex core location. **(D-F)** Phase field modeling demonstrating reversal of buckling pattern with applied field

In summary, this work presents two broad implications. First, it has been shown that SHG-based CD measurements using laboratory-based confocal optical measurements can be an effective probe of the chirality in such vortex arrays with a spatial resolution in the range of ~300 nm. In turn, these studies show that the vortex arrays assemble into domains of opposite chirality, identified by the change in the sign of the CD. The co-existence of regions of opposite circular dichroism (*i.e.*, chirality) points to an equilibrium between regions of opposite chirality and therefore to a pathway by which one region can be converted into the other with a thermodynamic field. Of greater importance is the observation that an electric field can switch the sign of the local chirality in a reversible, controlled and deterministic fashion. To the best of our knowledge, this is one of but a few instances where the chirality of a material can be manipulated with thermodynamic fields other than the chemical field. This work also points to a host of possibilities including probing the dynamics of such chiral switching, the role of the chiral-domain boundary and the microscopic structure of such domain boundaries.

**Materials and Methods**

<u>Sample Preparation</u>

The $[(PbTiO_3)_n/(SrTiO_3)_n]$ (n=16, 20) trilayers were synthesized on single-crystalline $DyScO_3$ (110) substrates via reflection high-energy electron diffraction (RHEED)-assisted pulsed-laser deposition (KrF laser). The $PbTiO_3$ and $SrTiO_3$ were grown at 610ºC in 100 mTorr oxygen pressure. For all materials, the laser fluence was 1.5 J/cm$^2$ with a repetition rate of 10 Hz. RHEED was used during the deposition to ensure the maintenance of a layer-by-layer growth mode for both the $PbTiO_3$ and $SrTiO_3$. The specular RHEED spot was used to monitor the RHEED oscillations. After deposition, the trilayers were annealed for 10 minutes in 50 Torr oxygen pressure to promote full oxidation and then cooled down to room temperature at that oxygen pressure.

<u>SHG CD Measurement</u>

SHG circular dichroism measurements were made in the reflection mode, using a Ti/Sapphire oscillator for excitation with ~100 fs pulses and center wavelength of 800 nm, a 78 MHz repetition rate, and an average power of 1 mW as shown in Supplementary Fig. 4. Pulses with linear polarization were sent through a ¼ wave plate (l/4) to generate LC or RC polarized excitation, and the light was then sent through a beam splitter (BS) and focused on the sample using an oil-immersion objective lens (OL, NA = 1.4). The back-scattered SHG signal was sent through a short pass filter (SP) and detected using a spectrometer (SpectraPro 500i, Princeton Instruments) with a charge-coupled device camera (CCD, ProEM+: 1600 eXcelon3, Princeton Instruments). Confocal scanning microscopy was used to create images of the SHG intensity generated with LC polarized excitation light and, subsequently, with RC polarized excitation light. These were used to obtain images by calculating the CD at each image pixel.

## Capacitance Measurements

The *in-plane* polarization vs. electric field measurements were performed using interdigitated electrodes (IDE) since there is no bottom electrode. The electrodes were at first patterned using a photolithographic lift off process with g-line photoresist, followed by Pt sputtering (100 nm) as the electrode. The photoresist was removed by sonicating with acetone, rinsed in deionized water, and blown dry with nitrogen. The permittivity of the film was determined from the measured capacitance using the partial capacitance model as described by Farnell[41]. In our case the length of the fingers was 500 μm, the distance between two fingers was 8 μm and the number of fingers was 40.

## TEM Measurements

The planar view TEM sample was mechanically polished using an Allied High Tech Multiprep at 0.5º wedge angle, and then ion-milled initially at 4 keV and then finished with 200 eV for final cleaning using a Gatan Precision ion polishing system. To image the local microstructure and detect the polarization distribution of the planar view TEM sample, atomic resolution STEM images were acquired on a spherical aberration (Cs) corrected FEI Titan 80–300 microscope operated at 300 kV with a point-to-point resolution of 50 pm at the National Center for Electron Microscopy, Lawrence Berkeley National Laboratory. Pairs of orthogonal scan images were used to correct the nonlinear drift distortions of the microscope in STEM mode. Using an established displacement vector-mapping algorithm[13,22] the local non-centrosymmetry of atomic columns in the $ABO_3$ lattice was probed to determine the polar structures.

## 4D STEM measurements

4D STEM measurement were carried out using a Gatan K3 direct detection camera located at the end of a Gatan Continuum imaging filter on a TEAM I microscope (aberration-corrected Thermo Fisher Scientific Titan 80–300). The microscope was operated at 300 kV with a probe current of 100 pA. The probe semi-angle used for the measurement was 2 mrad. Diffraction patterns were collected using a step size of 1 nm with 514x 399 scan positions. The K3 camera was used in full-frame electron counting mode with a binning of $4 \times 4$ pixels and camera length of 1.05 mm. The exposure time for each diffraction pattern was 47 ms.

Phase-field Simulations

In the phase-field modeling of the PbTiO$_3$/SrTiO$_3$ multilayer system, the evolution of the polarization was obtained by solving the time-dependent Ginzburg-Landau (TDGL) equations:

$$\frac{\partial P_i(\vec{r}, t)}{\partial t} = -L \frac{\delta F}{\delta P_i(\vec{r}, t)} (i = 1,2,3),$$

(3)

where $L$, $r$, and $t$ denote the kinetic coefficient, spatial position vectors, and time, respectively. The contributions to the total free energy F include the Landau bulk energy, elastic energy, electric (electrostatic) energy, and gradient energy, i.e.,

$$F = \int (f_{Landau} + f_{Elastic} + f_{Electric} + f_{Gradient}) dV$$

(4)

Expressions for each energy density can be found in the literature[42]. Due to the inhomogeneity of elastic constants in the [(SrTiO$_3$)$_9$/(PbTiO3)$_9$]$_n$ multilayer system, a spectral iterative perturbation method was employed to solve the mechanical equilibrium equation to obtain the stress field[43]. The pseudo-cubic lattice constants for PbTiO$_3$ and SrTiO$_3$ were taken as 3.9547 Å and 3.905 Å, respectively[43], while the anisotropic in-plane lattice constants for the substrate DyScO$_3$ is taken from literature to calculate the misfit strain[44]. Material constants for PbTiO$_3$ and SrTiO$_3$ used in the simulations are found from the literature and these include the Landau potentials, elastic constants, electrostrictive coefficients, background dielectric constants, and gradient energy coefficients[43]. Three-dimensional phase-field simulation of $[(\text{SrTiO}_3)_9/(\text{PbTiO}_3)_9]_n$ multilayer system, is done using discrete grids of $(120\Delta x) \times (120\Delta y) \times (350\Delta z)$ with $\Delta x = \Delta y = \Delta z = 0.4$ nm, where $\Delta x$, $\Delta y$, and $\Delta z$ are in real space. The thickness of the substrate, film, and air are $15\Delta$, $(18n + 9)\Delta z$, and $(326 - 9n)\Delta z$, respectively, where $n$ is the number of the confined PbTiO$_3$ layer. In the film, alternating 9 grids of PbTiO$_3$ layers and 9 grids of SrTiO$_3$ layers are incorporated to simulate the multilayer system. The periodicity of PbTiO$_3$ and SrTiO$_3$ layers effectively describes the experimental observation. Periodic boundary conditions are assumed in both the $x$ and $y$ directions, and a superposition method is used along the $z$ direction[45]. Random noise is used as the initial setup to simulate the thermal fluctuation during the annealing process.

<u>Second Principles Calculations</u>

The second-principles simulations were performed using the same methodology presented in previous works[39,46], as implemented in the SCALE-UP package[39,40]. The interatomic potentials, and the approach to simulate the interface, are the same as in Ref. [20]. We impose an epitaxial constraint assuming in-plane lattice constants of $a$=$b$=3.911 Å forming an angle of $\gamma$=90°. This corresponds to a small tensile epitaxial strain of +0.25% with respect the reference structure used in previous works (where $a$=$b$=3.901 Å) and mimics the mechanical boundary conditions imposed by the DyScO$_3$ substrate. This epitaxial condition favors the onset of an in-plane component of the polarization, that couples with the offset of the cores of the vortices.

For computational feasibility, we have focused on a simulation supercell made from a periodic repetition of $2n \times 1 \times 2n$ elemental perovskite unit cells, sufficiently large to simulate domains in the $n$=10 superlattice. At low temperatures the vortices do not vary along the axial $[1\bar{1}0]_o$ direction. Therefore, the simplification of taking only one unit cell along this direction does not affect the model while it speeds-up the calculations. For a given value of the electric field, we solved the models by running Monte Carlo simulated annealing down to very low temperatures, typically comprising 10,000 thermalization sweeps, followed by a conjugate gradient relaxation to find the ground state or metastable solutions. Local polarizations are computed within a linear approximation of the product of the Born effective charge tensor times the atomic displacements from the reference structure positions divided by the volume of the unit cell. From the complex polarization texture, we computed the helicity following equation 1. The curl of the local polarization that appears in the integrand is computed by a five-point finite difference central method. In order to improve the numerical stability of the derivatives we interpolated the polarization array via cubic splines to generate a fine-grained grid.

The experimentally observed samples correspond to the structure sketched in Fig. 1e, where the antiparallel axial components of the polarization in consecutive vortices coexist with an offset of the core of the vortices. This can be considered as a juxtaposition of the models sketched in Figs. 1c and 1d. As probed in the Supplementary Fig. 10, the switching of chirality in these structures is totally due to the switching of chirality of the model depicted in Fig. 1d. For computational feasibility, we focus on this model, saving computational resources since it does not require the presence of domains. For that we apply, on the one hand, a constant external field along the $[1\bar{1}0]_o$ direction to force the parallel orientation of the axial (Bloch[39,46]) component of the polarization at the core of neighboring polar vortices. On the other hand, another constant external field is applied along the $[110]_o$ direction to favor one of the out-of-plane domains (the one parallel to the external field) against the other. As shown in the Supplementary Section A, the final structure is chiral since it cannot be mapped by any combination of rotations and/or translations onto its three orthogonal reflections. To perform the hysteresis loop we started from a relaxed structure at zero electric field, that due to the tensile strain presented a given buckling of the vortex cores. Afterwards we used the relaxed structure as seed for the next calculation changing the value of the electric field. We used 13 values for the electric field along the $[001]_o$ direction, ranging from – 66.82 kV/cm to +66.82 kV/cm. This procedure was maintained during all the hysteresis loops.

**Acknowledgments:**

Office of Science, Basic Energy Sciences at the Department of Energy contract No. DE-AC02-05CH11231. (P.B., M.M., S.S. and R.R.)

University of California office of the President and the Ford Foundation. (L.C)

Laboratory Directed Research and Development Program of Lawrence Berkeley National Laboratory under U.S. Department of Energy Contract No. DE-AC02-05CH11231. (A.R)

The optical spectroscopy and electron microscopy work at the Molecular Foundry was supported by the Office of Science, Office of Basic Energy Sciences, of the U.S. Department of Energy under Contract No. DE-AC02-05CH11231.

Spanish Ministry of Science and Innovation through grant number PGC2018-096955-B-C41. (F.G.O, P.G.F., and J.J.)

Spanish Ministry of Universities through grant number FPU18/04661. (F.G.O.)

**Author contributions:**

R.R. and M.B.R. conceived the experiment.

M.A.M, M.R.M, L.C, and P.B. performed the SHG CD measurements.

M.A.M, A. R., and E. S. B. designed and set up the SHG CD measurements.

S-L.H. performed the HAADF-STEM and 4D-STEM imaging

S.S., B.H.S., and C.O. analyzed the 4D-STEM data

S-L.H. and P.B. performed PFM measurements.

C.T.N performed in-situ DF-TEM experiments.

S.D. prepared samples and performed capacitance measurements.

Z.H. performed phase field modeling.

F.G.O, P.G.F, and J.J conceived the theoretical model for the switching and performed the second principles calculations

S.W.L and G.v.d.L. developed theory for SHG-CD.

**Competing interests:** The authors report no competing interests.

**Data and materials availability:** All data are available in the main text or the supplementary materials.

Supplementary Materials

A.  <u>Different Models of Chirality</u>

Chiral behavior in three dimensions can arise in vortex structures as a consequence of a few symmetry-breaking pathways, which are summarized in Fig. 1 (main text). The first one is the coexistence of vortices with a polarization aligned along the normal of the plane containing the vortex. In PbTiO$_3$/SrTiO$_3$ superlattices, the driving force for the appearance of this axial component can be traced back to the tendency found in common 180º domain walls (DWs) of bulk PbTiO$_3$ to have a Bloch-like character at low temperatures, with a spontaneous electric polarization confined within the DW plane. The DW-DW interaction is so small than the energy split between the parallel and antiparallel configurations of the axial component at neighboring domain walls is negligible. A chiral configuration naturally occurs if an antiparallel axial polarization at the cores of neighboring clockwise/counterclockwise polar tubes condenses, as shown in Fig. 1c.

Even if the axial components align parallel, a second source of chirality can be found if the vortices have non-equal fraction of dipoles pointing along the $[110]_o$-direction (i.e., if the up and down domains are not exactly equal in size) superimposed on an offset of the vortex cores (Fig. 1d). These misalignments immediately yield a mismatch in the polarization pointing along the $[001]_o$-direction (left/right), which generates an excess of the clockwise/counterclockwise rotations. Therefore, for a given size of the domains, the structures adopt a chirality that is dependent on the direction of the mismatch.

The first mechanism described above gives very large values of helicity, since it combines a large value of the rotational component with the coupling with the axial component. Moreover, the contribution of each vortex tube sums in the same direction. However, the switching of chirality would be extremely costly: the two enantiomers would belong to the same topological class and can be transformed into each other in a continuous way. But during this process, some phases with head-to-head and tail-to-tail configurations will inevitably appear, making the path very unlikely, Fig. S3.

The second mechanism produces helicities that are typically two orders of magnitude smaller arising from an imbalance in the rotational component of the two vortices (Figure S2). But it has the advantage that it can be deterministically reversed in a fully controlled way by an external electric field that keeps track of the offset, and therefore the predominant left/right polarization direction, Fig. S1b. Due to the energetic degeneracy in the direction of the axial component, it is likely that left-handed and right-handed domains can coexist in the same sample, Fig. 1e (main text). In such a case, a chiral structure is expected assuming that there exists a mismatch between the axial components combined with the presence of an offset, where the high axial component of the polarization changes from the clockwise to the counterclockwise vortices in adjacent domains as sketched in Fig. 1e.

B.   Origin of chirality in the case of parallel axial components

Fig. S2

C.   Pathways to Switch Chirality with Antiparallel Bloch Components in the Domain Wall

When the axial component of the polarization points in opposite directions at the center of the clockwise and counterclockwise vortices, a chiral structure with a large value of the helicity is formed. Both the right- and left-hand enantiomers belong to the same topological class. Therefore, a continuous transition to transform one into the other might be envisaged. Here we propose three different paths, sketched in Fig. S3. The first one (top row in the central panel of Fig. S3) consists of a continuous 180° rotation of all the dipoles that form the vortices. However, there is one midpoint along the path where energetically cost head-to-head and tail-to-tail domains appear. Thus, although topologically allowed, such a transition implies a huge energy barrier, that makes it unlikely. The second mechanism consists in a continuous approximation of the vortex cores, as sketched in the central panel of Fig. S3. But again, the energy barrier to overcome is large: one of the domains along the $z$-direction increases its volume at the expense of the other, and this would translate into large depolarizing fields. Moreover, the Néel-components of the polarization would form head-to-head and tail-to-tail domains. The third path (bottom row in the central panel of Fig. S3) can be described as a homogenous reduction of the axial, Bloch component within the two vortex cores. At some point along the path, the axial component of the polarization would vanish, giving rise to an achiral structure. Beyond this point, a polarization at the center of the vortices opposite to the original one can be developed, changing the sign of the helicity. However, this procedure seems to be impractical, due to the difficulty of applying a spatially dependent external field that change its value in the length-scale of the separation of the vortex cores (around 8 nm).

D.  Helicity Computation of Chiral Model

Revisiting equation 1, we can observe that two ingredients are necessary to develop helicity. First, we need the curl of the polarization pattern to be non-zero. Second, this curl has to couple with the axial component of the polarization. As discussed in the main body of the manuscript, in the model of Fig. 1e we observe: first an offset between the center of the vortices along the $[110]_o$ direction (responsible for an asymmetry between the polarization components along the $[001]_o$ direction); second, a difference in the sizes of the up and down domains; and third, antiparallel axial components of the polarization at the center of the cores, being one of them slightly larger than the other. In such a model, there are two different sources of chirality. On the one hand, the existence of asymmetries between the UP/DOWN and the LEFT/RIGHT components of the polarization generates an imbalance between the curl of the clockwise and counterclockwise vortices, as shown in the Fig. S2. This is the driving force for the chiral behavior of the structure sketched in Fig. 1d. On the other hand, the anti-parallel axial components of the polarization contributes to the coupling between the curl and the final value of the helicity, responsible for the chirality of the structure discussed in Fig. 1c.[20]

Numerically we can assign relative values to both the curl and the axial component. Let us normalize the bare curl contribution to a value of 1 (positive if it is clockwise, and negative it is counter-clockwise). The offset of the center of the cores and the imbalance between the up/down domains induce an increase/decrease of this value by a given amount $\pm\delta$. For instance, in the leftmost tube of the front domain of the sketch of Fig. 1e, a configuration whose curl and domain structure are the same as the one in Fig. S3, the upper clockwise vortex has a curl of $(1-\delta)$, while the down counterclockwise vortex has a curl of $(-1-\delta)$. The fact that the curl is decreased in this case can be deduced from the fact that the negative magenta regions of the curl are more extended than the positive green region in the Fig. S2. The curls point in the axial direction.

Now, let us also normalize the maximum axial components of the polarization to a value of 1, represented by the dark red circle in the leftmost tube of Fig. 1e. Assuming that this axial component is slightly different in both directions (as schematized in Fig. 1e by the different intensity of the red/blue colors), the axial component in the neighboring vortex takes a value of $\left(-\frac{1}{a}\right)$, were $a$ is a parameter that accounts for the mismatch between components (if $a = 1$, then the two axial components are the same; if $a \to \infty$, one of the axial components vanishes).

Therefore, according to equation 1, the value of the helicity considering the two domains in Fig. 1e is as follows

$$\mathcal{H} \; = \; +1 \cdot (+1 - \delta) - \frac{1}{a}(-1 - \delta) - \frac{1}{a} \cdot (+1 - \delta) + 1(-1 - \delta) = -2\delta + \frac{2\delta}{a}$$

For any value of $a > 1$, i.e. assuming a slight difference in the axial component of the polarization, the helicity of the configuration in Fig. 1e will be negative.

Taking the mirror symmetry of the domain structure of Fig. 1e, as shown in Fig. 1f, the sense of the offset would be reverted, resulting on a change of the curl values that would amount to $(1 + \delta)$ for the clockwise vortices and $(-1 + \delta)$ for the counterclockwise. Computing again the helicity, this would result in a value of $+2\delta - \frac{2\delta}{a}$, the opposite as before, yielding to a change in the handedness.

E.   SHG CD Experimental Details

SHG-CD measurements were made as described in the methods section and depicted in Fig. S4a. To perform an SHG-CD measurement, the integrated intensity of the SHG signal at a single location on the sample was measured using RC polarized excitation light, producing a response with intensity $I_{RH}$. Next, the intensity was measured at the same location using LC polarized excitation to find $I_{LH}$. and the CD signal was calculated. Fig. S4b,c shows examples of SHG imaging with RC and LC excitation and Fig. S4d shows example spectra acquired in a RC domain (top) and an LC domain (bottom). Fig. S4e shows a 2D map of the SHG CD corresponding to the many integrated spectral pairs in parts **e** and **c** calculated using equation 2 in the main text. The characteristic length scale is shown by the line profile in Fig. S5f with a FFT of the image shown in the inset revealing domain elongation along the $[001]_o$ axis.

To further validate the SHG-CD measurements we performed a series of control experiments outlined in Fig. S5. First, we show an SHG-CD image (5c) along with the LC (5a) and RC (5b) images for comparison. Next, we measured SHG maps with 0° (5d) and 90°(5e) linearly polarized excitation, corresponding to polarizations along the $[001]_o$ and $[1\bar{1}0]_o$ sample axes, respectively. From these, we calculated the SHG linear dichroism (LD) shown in 6f using the equation $LD = \frac{I_{90°} - I_{0°}}{I_{90°} + I_{0°}}$. This reveals LD that is much weaker than the observed CD, confirming that the CD signal cannot simply be an artifact related to large LD anisotropy. In contrast to the chiral domains composed of polar vortices, we have measured SHG CD for the traditional ferroelectric *a*-domains which form in a (PTO)ₙ/(STO)ₙ trilayer heterostructure with $n = 8$ as an achiral control structure. We observe negligible CD signal for this sample, on the order of 10% of that observed in the $n = 20$ sample, as shown in (g)-(j), indicating that the CD signal is indeed related to polar domain formation. Finally, we confirmed that we are creating pure LC and RC light by comparing the normalized sum of the LC and RC excitation images (5j) to an image taken with linear excitation (5k). These images should look the same because linear polarized light is a superposition of LC and RC polarized light and thus excites both domain types equally. While some variation exists, these images agree to within the expectation. It is worth noting that the color maps for the calculated images shown in 6c, 5f, and 5j have been normalized to the same values so that their intensities can be directly compared.

## F.   Origins of Second Harmonic Generation

Second harmonic generation is a photonic exchange between the frequency components of the electromagnetic field. During the interaction with matter, two photons of lower frequency, $\omega$ ,are absorbed and one photon of $2\omega$ is created in a single quantum-mechanical process. Mathematically, an incident light source will interact with a medium's electronic structure, displacing the charged particles in the material during a process called polarization. Linear materials become polarized to a degree proportional to the intensity of the incident electric field

$$\mathbf{P} = \varepsilon_0 \chi^{(1)} \mathbf{E} \qquad (5)$$

where $\chi^{(1)} = n^2 - 1$ is the linear susceptibility tensor for refractive index $n$ which reflects the polarizability along each axis of the material. For very intense driving fields, however, nonlinearities in the material can be probed. The nonlinear polarization can be expressed as the Taylor expansion of the material's polarization in powers of the electric field with each component of $\mathbf{P}$ expressed as $P_k$ where $k = x, y, z$:

$$\mathbf{P} = \varepsilon_0 (\chi_{ik}^{(1)} \mathbf{E}_i + \chi_{ijk}^{(2)} \mathbf{E}_i \mathbf{E}_j + \chi_{ijlk}^{(3)} \mathbf{E}_i \mathbf{E}_j \mathbf{E}_l + \dots) \qquad (6)$$

where the coefficients of $\chi^{(n)}$ are tensors of increasing rank which correspond to the $n^{\text{th}}$ order process. These tensors describe the susceptibilities of the material along each axis and also coupling between material axes. The higher order processes are so weak that they were predicted but never observed until the advent the lasers in the 1960s[47].

Second harmonic generation arises from symmetry breaking in the second order susceptibility, which corresponds to a rank 3 tensor that describes a material's susceptibility along each axis. The second order polarizability goes as the electric field squared and for an incident electric field $E_i = \varepsilon_i e^{-i\omega t} + c.c$, it can be represented as

$$\mathbf{P}_k(NL) = \chi_{ijk}^{(2)}(\varepsilon_i \varepsilon_j e^{-i2\omega t} + \varepsilon_i^* \varepsilon_j^* e^{i2\omega t} + \varepsilon_i \varepsilon_j^* + \varepsilon_i^* \varepsilon_j) \quad (7)$$

From this equation, it can easily be shown that frequency doubling, or SHG, of the incident light is allowed.

Second-harmonic generation is sensitive to symmetry breaking, which can be seen by applying a spatial inversion operation to equation 7. The parity operation which takes $r \rightarrow -r$ changes the sign of both the polarization and electric fields. However, since the second order polarization goes as the square of the electric field, this term remains positive unless $\chi_{ijk}^{(2)}$, the susceptibility tensor, changes sign with the inversion. Only materials with susceptibility tensors that are sensitive to spatial inversion have a nonzero second order response. Physically this means that the material polarizes differently along different axes, and materials with this trait are commonly called noncentrosymmetric. This sensitivity to inversion symmetry makes second-harmonic generation a powerful tool for imaging crystal lattices with symmetry breaking, which often occurs at domain boundaries[30].

## G.  Circular Dichroism in Second Harmonic Generation

A SHG microscope can selectively observe the region in a sample where spatial inversion symmetry is broken. The attribute′s puissance has fueled a surge of applications of the technique to advance knowledge about nanostructures, molecular ordering, and structural organization in biological samples. Furthermore, the advent of free-electron lasers in the energy ranges from extreme ultraviolet to x-rays now allows to explore SHG effects involving core-level resonances.

A sophisticated set of theoretical tools at the disposal of exponents of ordinary NCD has been granted to users of NCD in the SHG response[35,48]. To begin with, ordinary dichroic signals use dyadic matrix-elements of the light-matter interaction found in the familiar Kramers-Heisenberg amplitude that is trivially extended to parity-odd E1-E2 or E1-M1 scattering events to cope with NCD[49]. Insight into the matrix elements is achieved by integrating out intermediate degrees of freedom, in the footsteps of Judd and Ofelt in their celebrated work on optical transition probabilities[50,51]. Specifically, a dyadic matrix element in the interaction is reduced to a spectrum of electronic multipoles routinely exploited in analyses and simulations of data gathered by other experimental techniques, including, NMR, EPR, Mössbauer effect, and resonant x-ray Bragg diffraction. Equivalent benefits accrue in an application of similar mathematical treatment to the NCD signal in the SHG response. In this case, a parity-odd E1′-E1-E1 process contains an NCD signal when angular momentum (**L**) and space (**R**) know inextricable knots which bind each to the other in the illuminated sample (E1 primary and E1′ secondary photon events)[35]. On the other hand, no NCD is allowed for the nominally weaker E1′-M1-M1 process.

A third-order perturbation theory accounts for the SHG response. The generic form of the NCD signal from the E1′-E1-E1 response is,

$$F(NCD) \, P_2 \sum_{\lambda, \lambda'} \{\langle g|x|\lambda\rangle\langle\lambda|y|\lambda'\rangle\langle\lambda'|z|g\rangle - \langle g|y|\lambda\rangle\langle\lambda|x|\lambda'\rangle\langle\lambda'|z|g\rangle\} \quad (8)$$

Here, coordinates are defined by a primary beam parallel to the z-axis and σ-polarization parallel to x-axis, and the electric dipole **R** = (x, y, z). The secondary wavevector, for E1′, is inclined to the z-axis and its polarization vector casts a shadow on the axis. $P_2$ is the pseudo-scalar Stokes parameter for circular polarization. Labels $\lambda$, $\lambda'$ delineate intermediate degrees of electronic freedom, while g denotes the ground-state.

The act in equation 8 of integrating out $\lambda$, $\lambda'$ creates atomic multipoles $\mathbf{U^K}$, where K is the rank. The expectation value of the dipole $\langle\mathbf{U^1}\rangle$ is the electric polarization in the electronic ground-state, for example. More generally, multipoles form observable signals that must be purely real, and the requirement restricts the composition of $\mathbf{U^K}$. For any process, such as E1′-E1-E1, there is a conjugate process that assures the response is purely real. Additional restrictions, or selection rules, flow from integrating out $\lambda$, $\lambda'$. Where upon, E1′-E1-E1 is no longer a simple product of three matrix elements of **R**. Rather, dipole matrix elements interlock in a very specific manner that reveals the nature of the SHG response, notably, selective replacement of **R** by **L**. One component of it remains after the fine sieve of photon polarization and the component in NCD is the quadrupole $\langle \mathbf{U}^2 \rangle$. Results for the SHG response are energy integrated signals. That is to say, from Eq. (8), $\langle U^2 \rangle \propto \{\mathbf{R} \otimes \mathbf{C}^2(\mathbf{R})\}^2$ is the total NCD signal available from a substance in a suitably designed measurement. Here, $\{. \otimes .\}^K$ denotes a standard tensor product of rank $K$, $\mathbf{C}^2(\mathbf{R})$ is a spatial spherical harmonic of rank 2 normalized such that $\mathbf{C}^1(\mathbf{R}) = \mathbf{R}$. The corresponding energy-integrated MCD signal is a correlation of orbital angular momentum and space, specifically, a sum of multipoles $\{\mathbf{C}^2(\mathbf{R}) \otimes \mathbf{L}\}^K$ with $K = 1$ and 3.[36]

## H. 4D-STEM

The 4D STEM data was analyzed using py4dstem software package[52]. Prior to determining the polarization from the diffraction patterns, the diffraction disks were corrected for diffraction shift and scan coil rotation in order to deterministically locate the $[001]_o$ and $[1\bar{1}0]_o$ directions. After all the calibrations, Bragg disks were detected on diffraction pattern at every scan position. Then two Friedel pairs, Fig. S6a, in the $[1\bar{1}0]_o$ direction (green and yellow circle) and $[001]_o$ direction (red and blue circle) were selected to generate the respective virtual images, Fig. S6d-e. The polarization was determined by subtracting the normalized intensity difference between virtual images formed by the red and blue circles for lateral or $[001]_o$ polarization and green and yellow circles for axial polarization $[1\bar{1}0]_o$.

## I. Electric Field Dependent Measurements

For electric field measurements interdigitated electrodes were deposited on the sample surface ( Fig. S7a). These electrodes create constant electric field with alternating sign across neighboring interdigitated regions.

The phase change between LC and RC chirality was characterized by measuring its hysteresis. This was done as described in the main text by recording the SHG CD for applied voltages starting at +60 V, decreasing in increments of 10 V to -60 V, and reversing the process for increasing voltages back to +60 V. This revealed clear hysteresis behavior as shown in Fig. S8a. Furthermore, this behavior was observed to follow the inverted pattern shown in Fig. S8b as expected for regions of opposite applied field. These hysteresis measurements were repeated at several sample locations with good reproducibility as shown by the data in Fig. S8 for regions of both field polarities.

In-plane capacitance measurements using similar devices to those used for Fig. S7a, yield the polarization vs. electric field data presented in Fig S9, in red. The large dielectric contribution was subtracted when presented in Fig. 4e. A similar procedure was followed for the polarization calculated from second principles (in black), where the lesser dielectric contribution was subtracted out when presented in Fig. 4e.

J.   Second Principles Simulation

Fig. S10

K.  <u>In-Situ DF-TEM Studies</u>

By changing the reflection conditions used in DF-TEM, we are able to probe the orientation of domains within the vortex phase. Focusing on the $[110]_o$ family of reflections, gives rise to contrast along the c-axis. This allows for differentiating between UP/DOWN domains, as seen in Fig. 4a-c and Fig. S11. Using the $[001]_o$, allows for differentiating between domains with opposing $[001]_o$ polarization, Fig. S11b. The combination of the two allows for the assignment of vectoral polarization depending on the intensity seen in DF-TEM.

Application of a tip on the surface of the cross-sectioned sample allows for probing of both in-plane and out-of-plane polarization. Fig. S12 illustrates that depending on the location of the center of the tip relative to the sample, there will be an increased in-plane field across the sample the further away from the tip center one images. The attached Movie S1 shows the transformation of the vortices from both a majority out-of-plane and majority in-plane polarization.

DF-TEM data for intermediate applied voltages throughout the chiral phase transition is provided in Fig. S13. Images were taken at increments of ~1.8V, with the majority of the field along the $[001]_o$ in-plane direction. The first transition is seen going from 3.7 to 5.6V, where the net in-plane polarization switching causing a reversal in the vortex buckling. As the voltage increases, the system approached a pure ferroelectric phase seen at 15.0V. This further demonstrates that the majority of the field lies in-plane for this location. A video of the transition has also been provided, which demonstrates reversibility of the chiral phase transition.

L.  <u>Chirality in Ferroic Materials</u>

Chirality is a widespread phenomenon in nature that indicates symmetry breaking in which an object's mirror image cannot be superimposed onto the original object. The chirality of a particle can have a profound effect on its behavior. For example, while all glucose molecules taste sweet, the human body can only metabolize naturally occurring glucose molecules, which have a right-handed chirality. On the other hand, glucose engineered by humans to have a left-handed chirality cannot be metabolized and is commonly used as a calorie-free sweetener.

Chiral phenomena in ferroic materials such as chiral magnetic domains[53] and chirality in nanoscale magnets[54] have been recognized as a promising path to extending Moore's Law for decreasing the spatial extent and energy consumption of information storage and processing devices using spintronics. In these materials, the spin-orbital interaction, called the Dzyaloshinskii-Moriya interaction, twists the magnetization to induce topological defects such as chiral magnetic, and topological chiral domain walls[55,56]. Additionally, the ability to tune the chirality of Bloch-type and Néel-type domain walls has been demonstrated[6,57,58] as an important step toward the control necessary to read and write spin-based bits.

However, domains in ferroelectric materials are usually separated by domain walls with uniform polarization and no significant chirality[59]. Complex ferroelectric domain walls such as Bloch-type and Néel- type domain walls are associated with local polarization rotation and offer a potential path to discovering chirality in ferroelectric materials[60,61]. However, topological defects such as ferroelectric vortex structures are generally considered the most promising candidates for applications in technology due to their ability to support clockwise (CW) and counterclockwise (CCW) rotating polarizations. For ferroelectric materials with CW and CCW polarization rotations in addition to an axial polarization component, it has been shown theoretically that the 3D vortex state can possess chirality with different handedness[62].

In recent studies, polar vortices in ferroelectric superlattices were created[13,63] by tuning the interactions between the electrostatic and elastic boundary conditions through the layer thickness, substrate, and choice of materials. In these materials, polar vortices form in ordered arrays with continuous rotation of the polarization and coexist with ferroelectric a-domains[22]. A significant discovery recently showed that polar vortices in $PbTiO_3/SrTiO_3$ superlattices exhibit strong circular dichroism when circularly polarized light is used in resonant soft X-ray diffraction, indicating that these vortices may have a chiral structure[20].

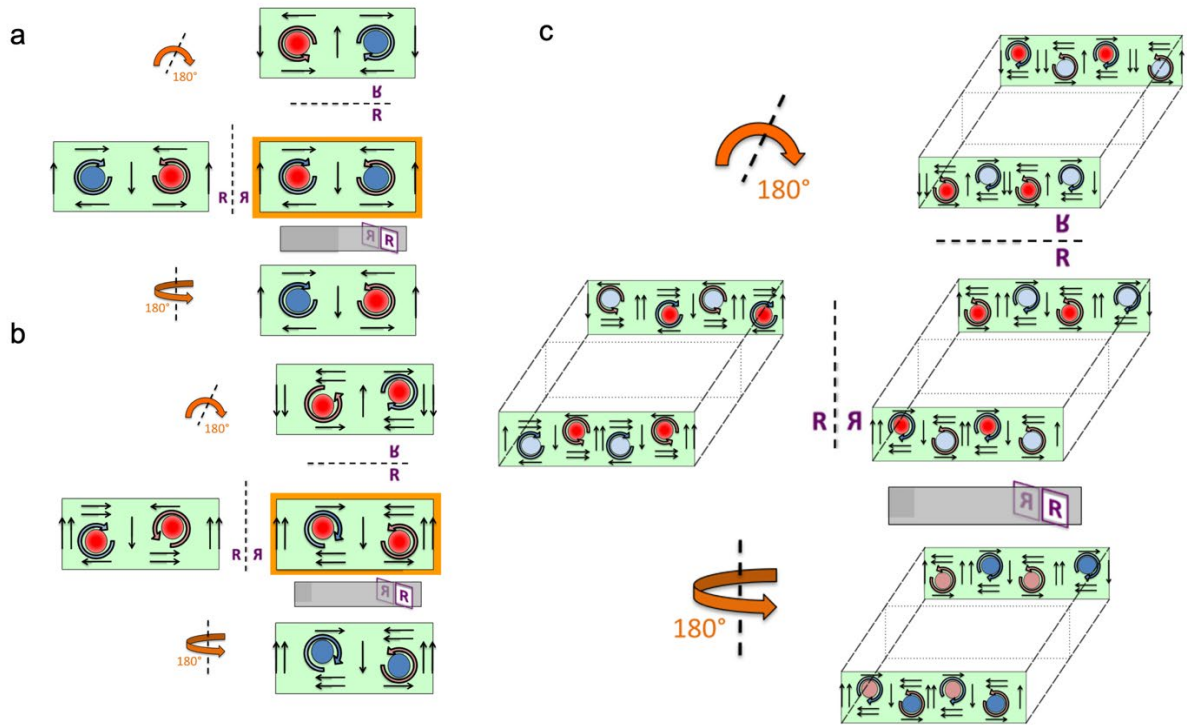

**Fig. S1:** Key features of the chiral character for the different models of Fig. 1 of the main text are schematically captured. Straight arrows point along the direction of the local polarization. Clockwise (counterclockwise) vortices are represented by blue (orange) curled arrows. Filled circles represent the direction of the axial component of the polarization, with light colors representing lower values in magnitude. Three orthogonal reflections are examined for each of the models shown in Fig. 1, which are perpendicular to $[001]_o$ (left), $[110]_o$ (upper), and $[1\bar{1}0]_o$ (bottom). Reflections can be mapped onto each other but not to the original one by means of rotations and translations, showing that they are chiral enantiomers.

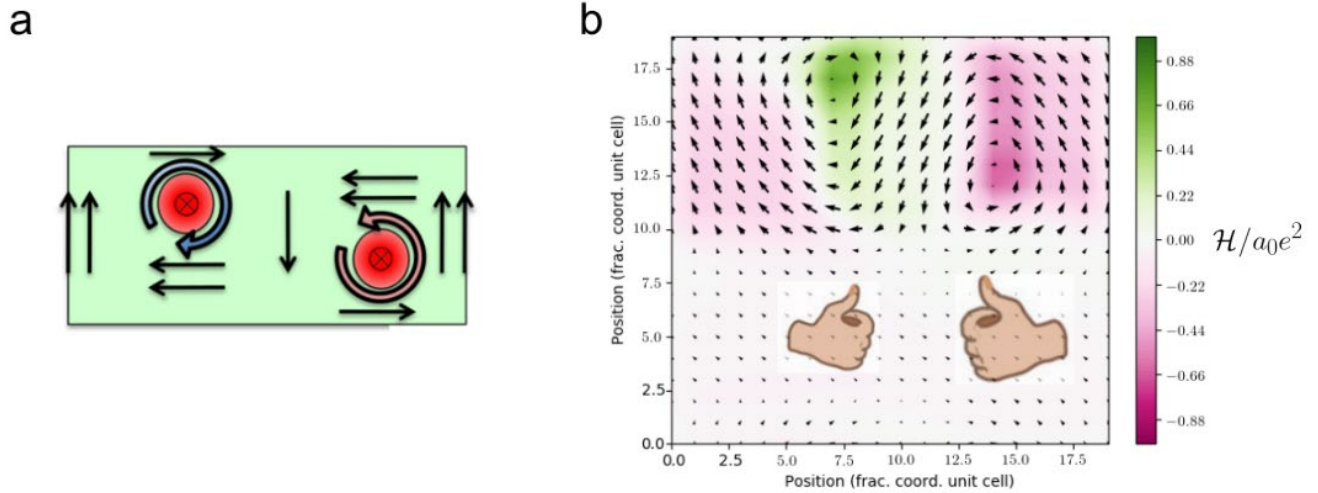

**Fig. S2: Origin of the chirality for the vortices with a parallel axial component of the polarization.** **(A)** Sketch of the non-trivial topological texture shown in Fig. 1d. An offset between the center of the vortices is observed along the [110]$_o$-direction, together with a different size of the up/down domains. The offset leads to a mismatch between the polarization pointing along the [001]$_o$ direction (in the cartoon, the polarization pointing to the left is predominant). The axial component of the polarization is parallel at the center of the two vortices (filled red circles). **(B)** Second-principles results for the same configuration. The arrows represent the in-plane component of the polarization. The colors represent the helicity density (integrand of equation 1), with a scale indicated by the color of the side bar. The handedness of the two vortices is opposite, sketched by the hands in the cartoon. Nevertheless, the combination of a majority of the left-pointing polarization (produced by the offset) with the majority of up polarization yields to the fact that the regions with a negative sign of the helicity (magenta regions) are dominant, making the whole system chiral, as highlighted with the different size of the two hands. In the absence of an offset between the center of the two vortices, there would be no net polarization along [001]$_o$ direction and the previous effect vanishes. If the offset changes the sign, keeping constant the sizes of the up and down domains, then the whole effect reverts its sign, making the right-hand (green regions) predominant.

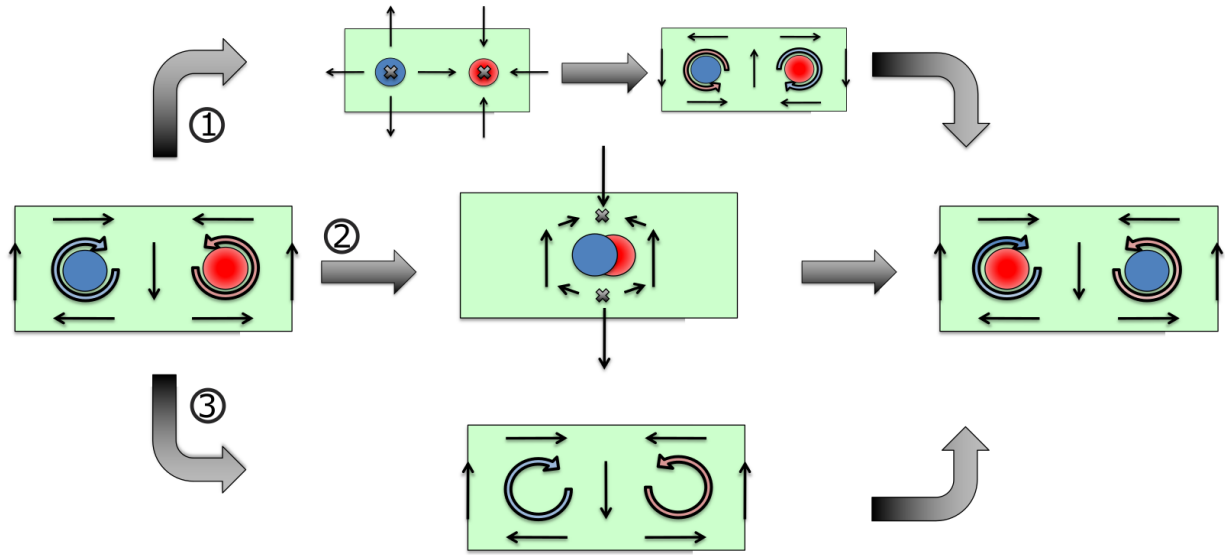

**Fig. S3:** Different paths to switch the helicity of the vortices when the chirality is due to the opposite direction of the axial polarization at the cores of neighbor vortices. Straight arrows represent the direction of the local polarization in the *xz*-plane, while the curled arrows indicate the sense of rotation of the vortices. Red and Blue circles represent different directions of the axial component of the polarization. The crosses are the points where head-to-head and tail-to-tail domains are formed.

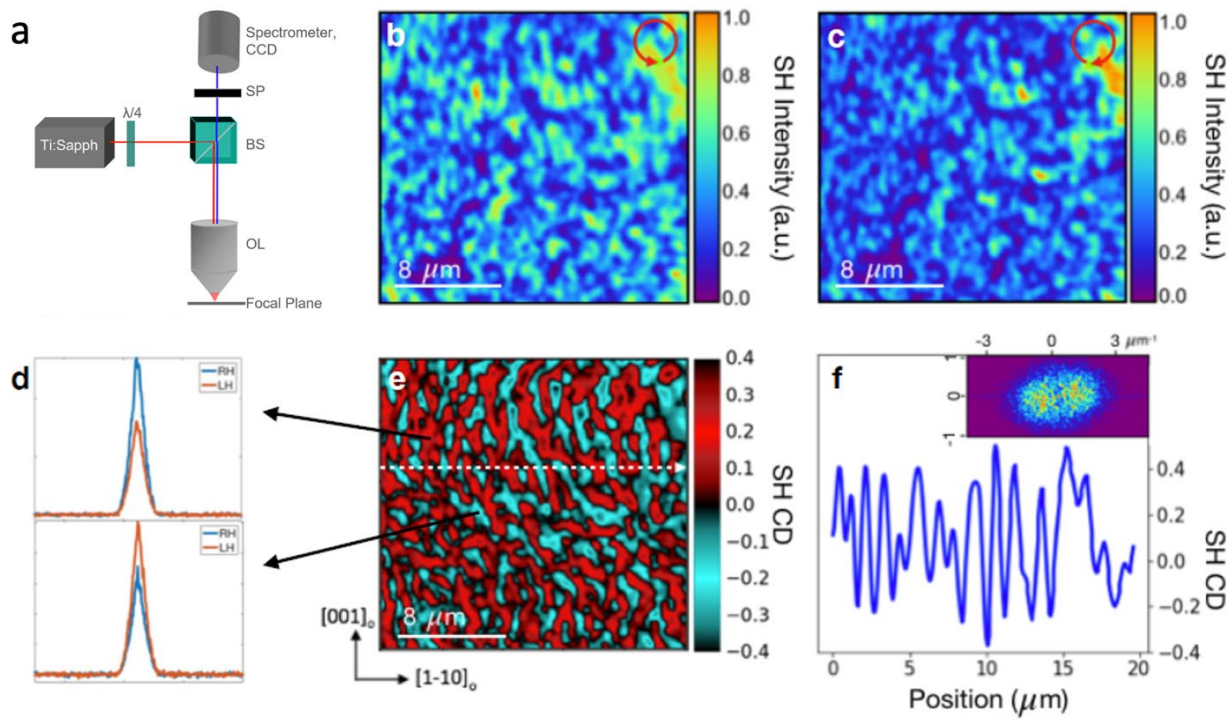

**Fig. S4: (A)** Experimental layout for the SHG-CD measurements. **(B,C)** Examples of SHG images acquired with RC/LC excitation. **(D)** Examples of SHG spectra acquired at a single point in a RC dominant domain (top) and a LC dominant domain (bottom). **(E)** SHG CD calculated from the images shown in (B) and (C). **(F)** Line profile along the dotted line shown in (E) and FFT of the SHG CD showing domain elongation along the $[001]_o$ axis (inset).

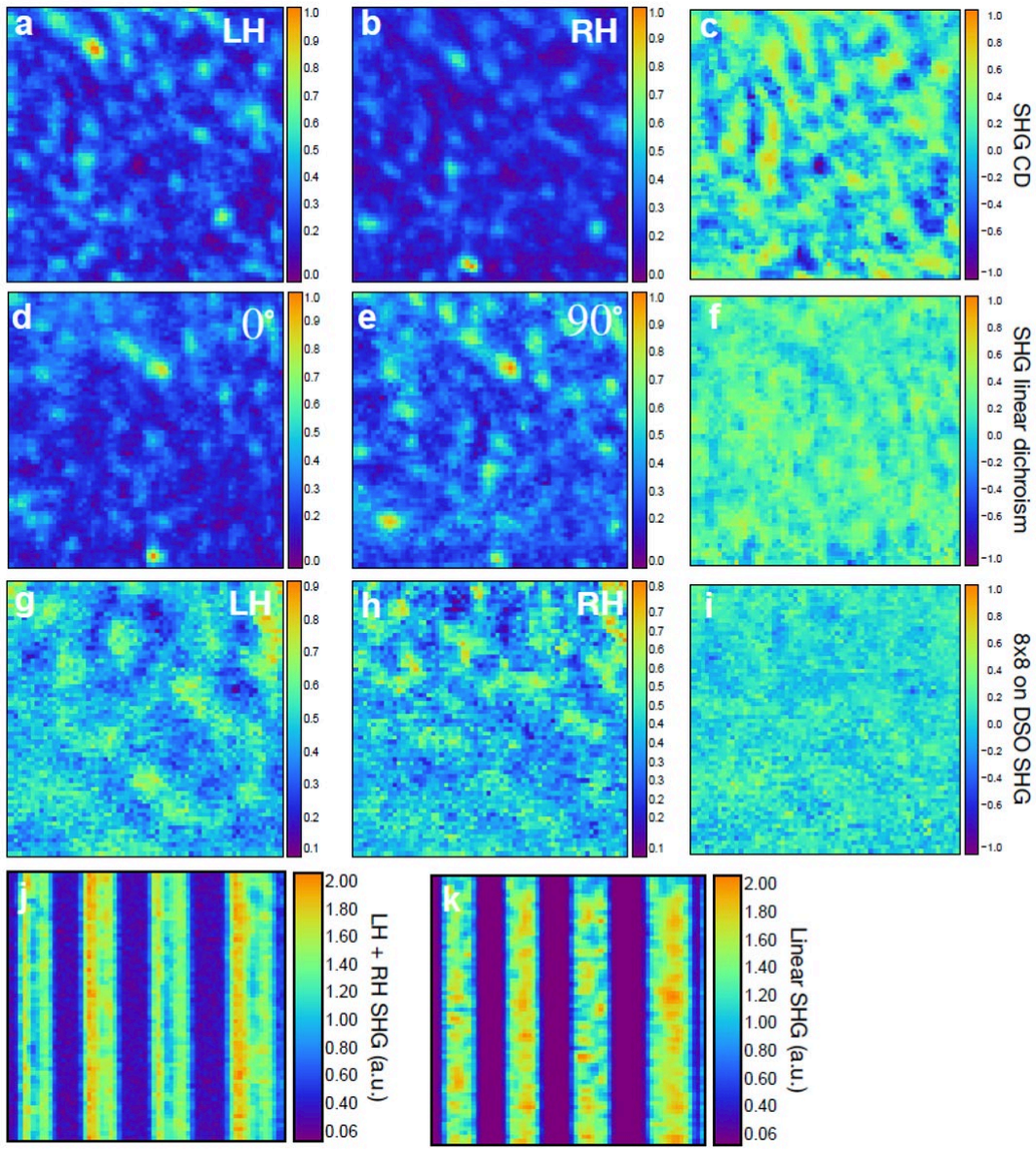

Fig. S5: **(A-C)** LC and RC SHG images and calculated CD for a normal sample region. d-f, Linear dichroism for the same region. **(G-I)** SHG CD for a ferroelectric trilayer structure with no vortex formation. **(J-K)** Sum of LC and RC images, and linear SHG showing that imaging with linear excitation excites both LC and RC modes.

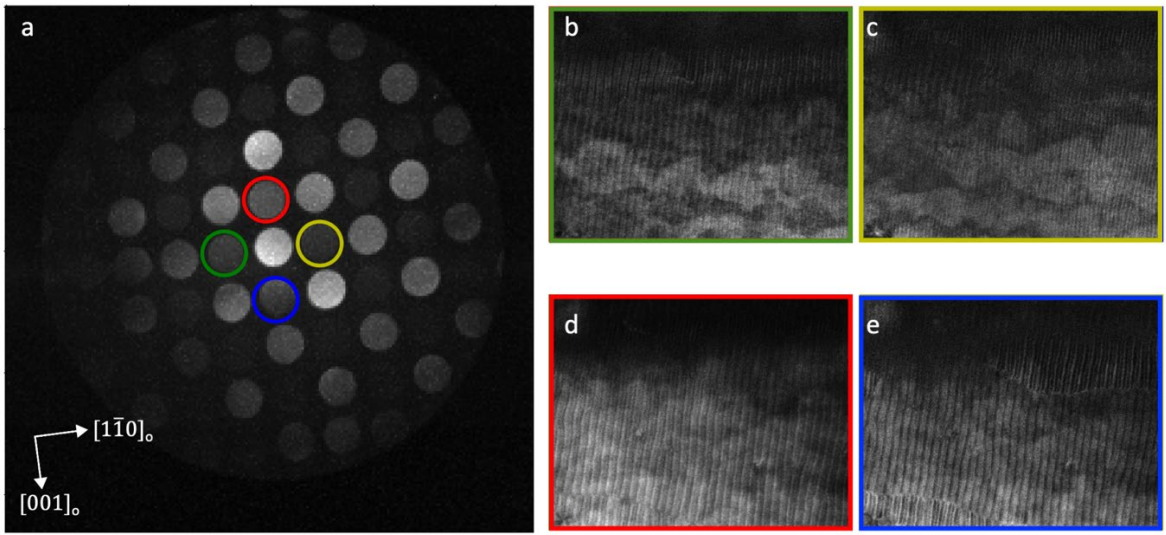

**Fig. S6: (A)** Normalized CBED pattern extracted from entire region in Figure 3a. **(B-E)** Virtual images that are formed by selected the different diffraction disks along $[001]_o$ (red/blue) and $[1\bar{1}0]_o$ (green/yellow) directions.

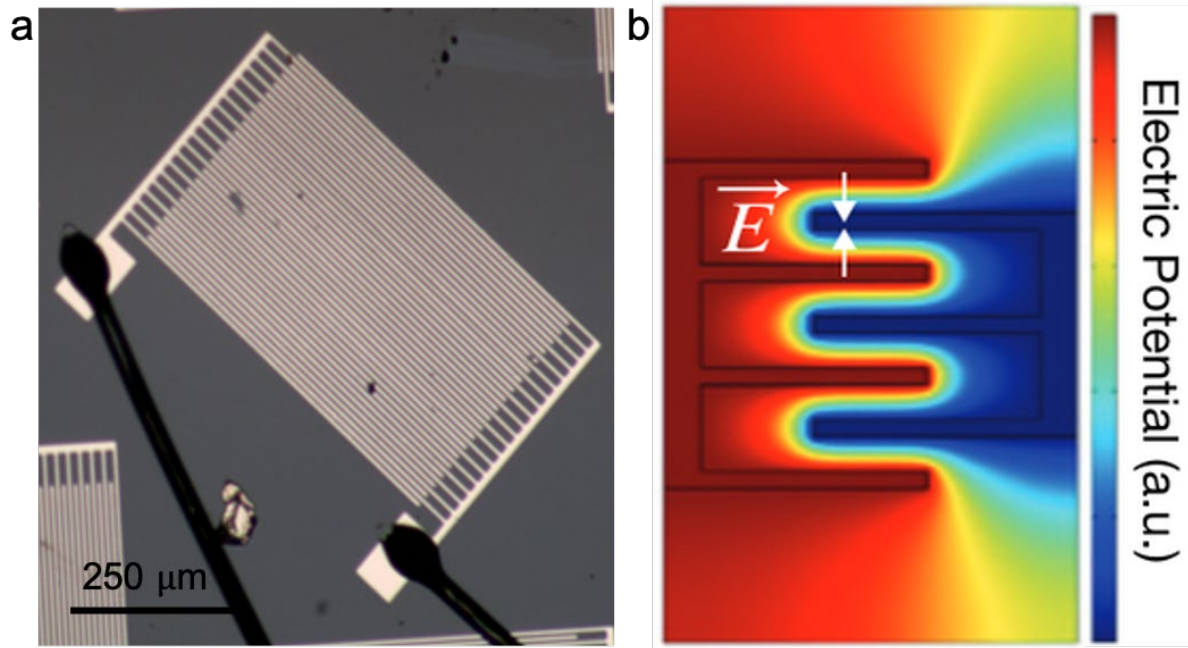

**Fig. S7: (A)** Image of representative IDE device used. **(B)** Calculated electric potential for and IDE device, demonstrating opposite electric field across neighboring fingers.

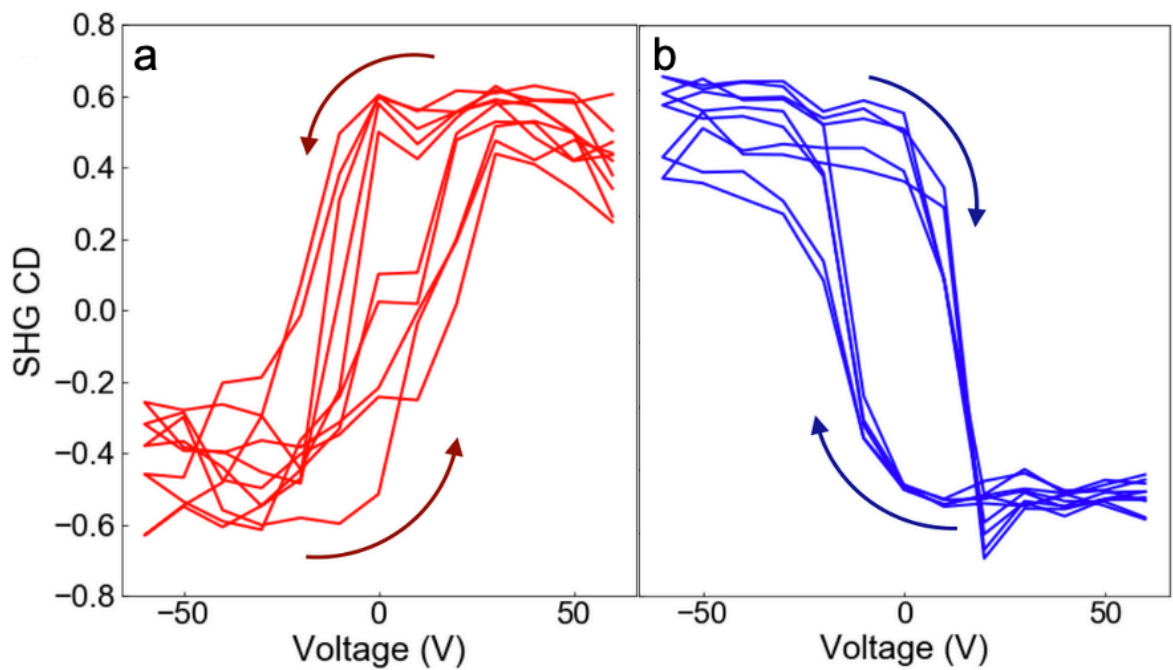

**Fig. S8: (A-B)** Hysteresis repeated in five different RC and LC oriented chiral domains, respectively, which show the expected mirrored behavior in the CD signal.

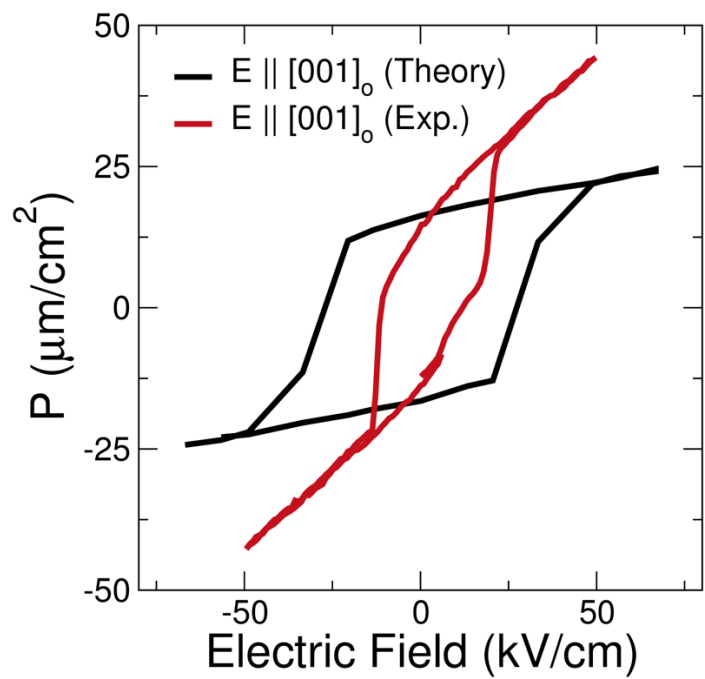

**Fig. S9:** Raw data for polarization vs. electrical field measurements performed with the electric field along the [001]$_o$ axis (red). Polarization response from second principles is seen in black.

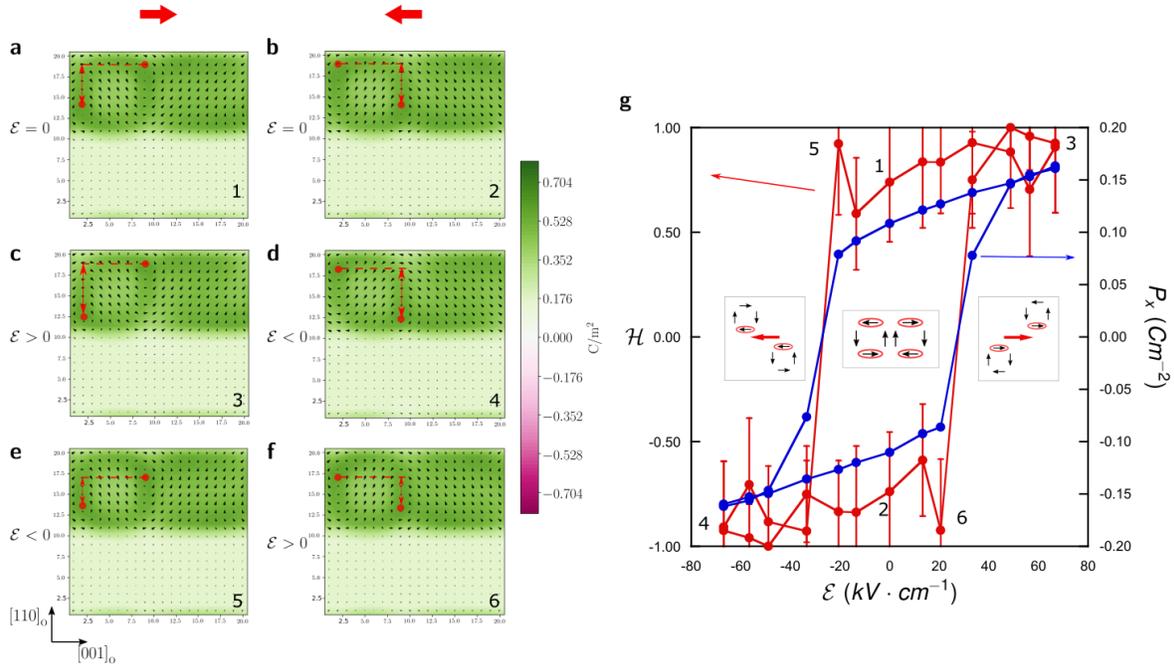

**Fig. S10: (A-B)** Enantiomer configurations of vortices with opposite buckling. Filled red points indicate location of vortex cores. Large red arrows at the top indicate direction of net polarization along $[001]_o$ induced by the offset. **(C-D)** Displacement of the vortices when the field is applied parallel to the net in-plane polarization. The vertical separation between the vortex cores is enhanced. **(E-F)** Displacement of the vortices when the field is applied anti-parallel to net in-plane polarization previous to the coercive field. The vertical distance between the vortex cores is progressively reduced. **(G)** Helicity (red; left axis) and polarization (blue; right axis) for model system as a voltage is applied along $[001]_o$. Numbering of points relates numerical values of the helicity to polarization patterns presented in (A-F). Insets schematize the direct coupling between the sense of the buckling and the $[001]_o$ component of the polarization.

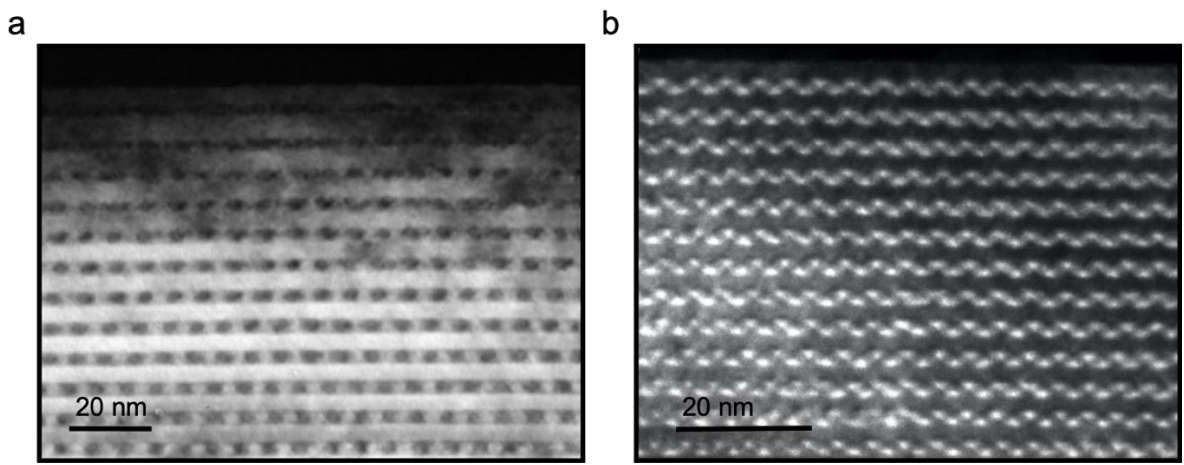

**Fig. S11: (A)** DF-TEM image taken using the $[220]_o$ reflection, giving rise to contrast from domains with opposing displacement along the *c*-axis. **(B)** DF-TEM image taken using $[008]_o$ reflection, giving rise to contrast from domains with opposing displacement along the *a*-axis

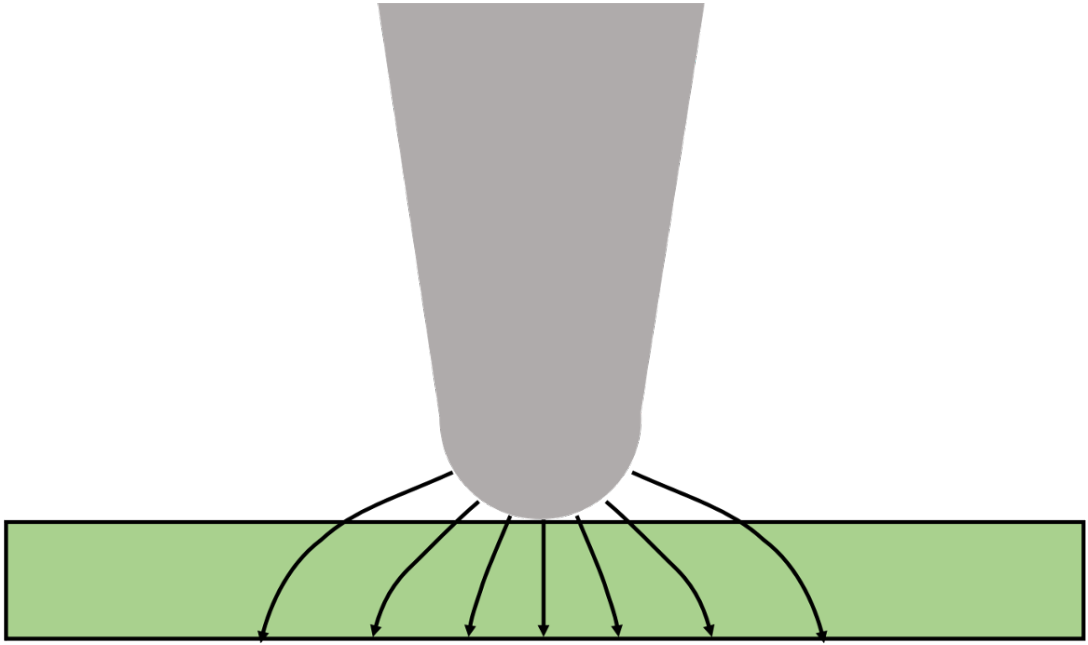

**Fig. S12:** Schematic of electric field applied through a tip. Demonstrates mixture of in-plane and out-of-plane fields.

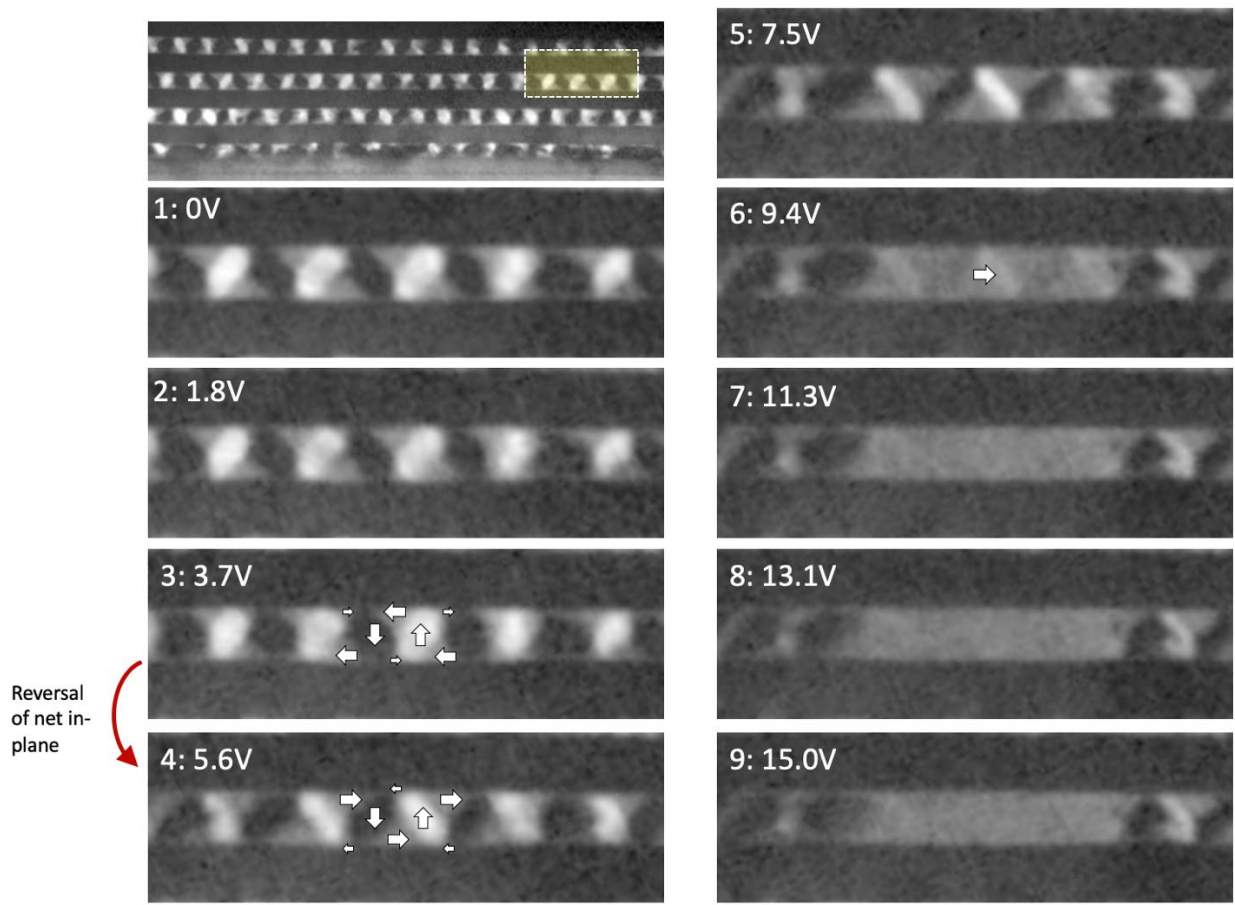

**Fig. S13: DF-TEM** imaging of chiral phase transition. DF-TEM images for increasing applied voltages showing microscopic restructuring across the in-plane reversal of polarization and transition to pure ferroelectric state at higher voltages